\documentclass[11pt,a4paper]{article}
\pdfoutput=1

\usepackage{epsf}

\usepackage{epsfig,graphics}
\usepackage{graphicx}
\usepackage{epsfig}
\usepackage{subfigure}
\usepackage{tabularx} 
\usepackage{rotate}	
\usepackage{slashed}
\usepackage{color}

\usepackage{amsmath}
\usepackage{amsfonts}
\usepackage{amssymb}
\usepackage{graphicx}
\usepackage{cite}

\newcommand{\bmat}{\left(\begin{array}}
\newcommand{\emat}{\end{array}\right)}

\def\yzero{\smash{\hbox{$y\kern-4pt\raise1pt\hbox{${}^\circ$}$}}}

\def\beq{\begin{equation}}
\def\eeq{\end{equation}}
\def\beqa{\begin{eqnarray}}
\def\eeqa{\end{eqnarray}}

\def\-{\hphantom{-}}

\def\s2{\frac{1}{\sqrt2}}

\def\beq{\begin{equation}}
\def\eeq{\end{equation}}
\def\beqa{\begin{eqnarray}}
\def\eeqa{\end{eqnarray}}

\def\IF{\relax{\rm I\kern-.18em F}}
\def\II{\relax{\rm I\kern-.18em I}}
\def\IP{\relax{\rm I\kern-.18em P}}
\def\IC{\relax\hbox{\kern.25em$\inbar\kern-.3em{\rm C}$}}
\def\IR{\relax{\rm I\kern-.18em R}}

\def\Dsl{\,\raise.15ex\hbox{/}\mkern-13.5mu D} 
\def\IZ{Z\kern-.4em  Z}

\def\-{\hphantom{-}}

\def\s2{\frac{1}{\sqrt2}}

\def\IC{\mathbb{C}}

\def\IF{\relax{\rm I\kern-.18em F}}
\def\II{\relax{\rm I\kern-.18em I}}
\def\IP{\relax{\rm I\kern-.18em P}}
\def\IC{\relax\hbox{\kern.25em$\inbar\kern-.3em{\rm C}$}}
\def\IR{\relax{\rm I\kern-.18em R}}

\newcommand{\be}{\begin{equation}} \newcommand{\ee}{\end{equation}}
\newcommand{\bea}{\begin{eqnarray}} \newcommand{\eea}{\end{eqnarray}}
\newcommand{\beann}{\begin{eqnarray*}}  \newcommand{\eeann}{\end{eqnarray*}}
\newcommand{\bfig}{\begin{figure}} \newcommand{\efig}{\end{figure}}
\newcommand{\ba}{\begin{array}} \newcommand{\ea}{\end{array}}
\newcommand{\bcen}{\begin{center}} \newcommand{\ecen}{\end{center}}
\newcommand{\btab}{\begin{tabular}} \newcommand{\etab}{\end{tabular}}

\newcommand{\matt}{\left ( \begin{array}{ccc}}
\newcommand{\ematt}{\end{array} \right )} \newcommand{\matf}{\left
(\begin{array}{cccc}}

\catcode`\@=11   
\newdimen\@rotdimen
\newbox\@rotbox  

\def\@vspec#1{\special{ps:#1}}
\def\@rotstart#1{\@vspec{gsave currentpoint currentpoint translate
   #1 neg exch neg exch translate}}
\def\@rotfinish{\@vspec{currentpoint grestore moveto}}
\def\@rotr#1{\@rotdimen=\ht#1\advance\@rotdimen by\dp#1%
   \hbox to\@rotdimen{\hskip\ht#1\vbox to\wd#1{\@rotstart{90 rotate}%
   \box#1\vss}\hss}\@rotfinish}
\def\@rotl#1{\@rotdimen=\ht#1\advance\@rotdimen by\dp#1%
   \hbox to\@rotdimen{\vbox to\wd#1{\vskip\wd#1\@rotstart{270 rotate}%
   \box#1\vss}\hss}\@rotfinish}%
\def\@rotu#1{\@rotdimen=\ht#1\advance\@rotdimen by\dp#1%
   \hbox to\wd#1{\hskip\wd#1\vbox to\@rotdimen{\vskip\@rotdimen
   \@rotstart{-1 dup scale}\box#1\vss}\hss}\@rotfinish}%
\def\@rotf#1{\hbox to\wd#1{\hskip\wd#1\@rotstart{-1 1 scale}%
   \box#1\hss}\@rotfinish}%
\def\rotate{\@ifnextchar[{\@rotate}{\@rotate[l]}}
\def\@rotate[#1]#2{\setbox\@rotbox=\hbox{#2}\@nameuse{@rot#1}\@rotbox}

\catcode`\@=12
\begin{document}

\makeatletter
\@addtoreset{equation}{section}
\makeatother
\renewcommand{\theequation}{\thesection.\arabic{equation}}


\pagestyle{empty}
\rightline{ DESY-16-224}
\rightline{IFT-UAM/CSIC-16-128}
\vspace{1.5cm}

\begin{center}
\LARGE{Inflation with a graceful exit in a random landscape} 
\\[13mm]
  \large{F. G. Pedro$^{1}$,  A. Westphal$^{2}$\\[6mm]}
\small{
${}^1$  Departamento de F\'{\i}sica Te\'orica
and Instituto de F\'{\i}sica Te\'orica UAM/CSIC,\\[-0.3em]
Universidad Aut\'onoma de Madrid,
Cantoblanco, 28049 Madrid, Spain\\
${}^2$ Deutsches Elektronen-Synchrotron DESY, Theory Group, D-22603 Hamburg, Germany\\[8mm]}
\end{center}

\begin{abstract}
We develop a stochastic description of small-field inflationary histories with a graceful exit in a  random potential  whose Hessian is a Gaussian random matrix as a model of the unstructured part of the string landscape. The dynamical evolution  in such a random potential from a small-field inflation region towards a viable late-time de Sitter (dS) minimum maps to the dynamics of Dyson Brownian motion describing the relaxation of non-equilibrium eigenvalue spectra in random matrix theory. We analytically compute the relaxation probability in a saddle point approximation of the partition function of the eigenvalue distribution of the Wigner ensemble describing the mass matrices of the critical points. When applied to small-field inflation in the landscape, this leads to an exponentially strong bias against small-field ranges and an upper bound $N\ll 10$ on the number of light fields $N$ participating during inflation  from the non-observation of negative spatial curvature.
\end{abstract}
\newpage

\setcounter{page}{1}
\pagestyle{plain}
\renewcommand{\thefootnote}{\arabic{footnote}}
\setcounter{footnote}{0}

\tableofcontents
\vspace{1cm}

\section{Introduction}

Cosmological inflation is the main contender for the description of the very early universe prior to the conventional hot big bang epoch. It has strong empirical support from various cosmological probes such as  e.g.  the cosmic microwave background (CMB) precision data, type Ia supernovae, and baryon acoustic oscillations (BAO). However, inflation in general is more of a paradigm, as the detailed microscopic model of inflation is unknown, with the available cosmological data allowing for large classes of inflationary scalar potentials. Moreover, at the theory level, inflation is sensitive to quantum gravity effects, and hence to the short distance (UV) completion of quantum mechanics and general relativity~\cite{Baumann:2014nda}. It is this property which motivates a study of UV completions of inflation in string theory as one of our best candidates for a theory of quantum gravity.

String theory is described at the worldsheet level by a two-dimensional conformal field theory (CFT). A large class of consistent effectively four-dimensional solutions to string theory, called `string vacua', arises by compactifying the six extra space dimensions arising from the excess central charge of the worldsheet CFT. These string compactifications typically come with a plethora of moduli scalar fields, parametrizing deformations of extra-dimensional manifold, and axionic pseudo-scalar fields from higher-dimensional gauge potentials.

Lacking observational evidence for their existence, these moduli are in need of stabilization in order to acquire large masses. A combination of tree-level sources such as quantized $p$-form field strengths (fluxes), perturbative string quantum effects, and non-perturbative effects such as D-branes and instantons serve to stabilize the moduli in a large discretum of meta-stable 4D string vacua, some of which can be cosmologically viable de Sitter (dS) vacua~\cite{Baumann:2014nda}. 

Among the many fields of this high-dimensional scalar potential `landscape', inflation can arise either by the `accident' via occurrence of a narrow slow-roll flat region in the scalar potential, or as a long large-field valley due to underlying structures and/or symmetries of a subsector of landscape. Examples for the latter large-field high-scale inflation models arise e.g. from the approximate shift symmetry of axion inflation models, or the asymptotic exponential series of certain volume moduli inflation models, for a recent review see~\cite{Baumann:2014nda}.

In this paper we study the part of the landscape without long-range structures. In this case we can approximate inflation as arising at random by local cancellations among terms in a random potential thereby producing a narrow slow-roll flat region supporting small-field inflation. One example for such sectors of the landscape is the scalar potential for the $h^{2,1}\gg 1$ complex structure moduli of a generic non-trivial Calabi-Yau compactification of string theory (at least, away from the limit of large complex structure~\cite{Brodie:2015kza}).

Previous work~\cite{Marsh:2011aa,Freivogel:2016kxc} has studied the probability of viable local dS minima in this context using the fact that the statistics of critical points and the eigenvalue distribution of their mass matrices (Hessians) in a random potential are well described by the statistics of sets of large Gaussian random matrices. Moreover, random matrix theory (RMT) has been applied in~\cite{Marsh:2013qca} in a reconstruction of a random potential along the path of steepest decent starting from a local critical point with a given mass matrix. This local reconstruction of the random scalar potential along the inflationary path rests on the description of the eigenvalue distribution of Hessian of critical points in Gaussian random potentials and its stochastic variation along random paths in field space by Dyson Brownian motion (DBM)~\cite{Uhlenbeck:1930zz,Dyson:1962}. The eigenvalues of an ensemble of Gaussian random matrices describing the critical point Hessians behave like a 1D gas of electrically charged particles with logarithmic mutually repulsive potential in a common quadratic confining potential. This picture allows for an intuitive understanding of the behaviour of the eigenvalue spectrum of the Hessians along trajectories in field space, including the effect of `eigenvalue repulsion'. Since eigenvalues tend to repel each other, moving along such a path in field space tends to rapidly generate strongly tachyonic directions in the Hessian. This is the reason why both local dS minima and inflationary small-field critical points are exponentially rare in such structure-less sectors of the string landscape.

In this paper we apply the description of the evolution of the eigenvalue distribution of critical point Hessians in a Gaussian random landscape via DBM to the question of finding a 'graceful exit' from  the inflationary regime of random inflationary small-field critical points. A 'graceful exit' from inflation describes the requirement of finding a local, cosmologically viable dS minimum after rolling away from some inflationary critical point in the landscape. We compute the corresponding probability of a graceful exit using the DBM process over multiple correlation lengths of the underlying random potential. We do this first by numerically integrating the discretized Dyson Brownian motion equations~\cite{Dyson:1962,Marsh:2013qca}.

Then we apply the description of the eigenvalue distribution via the 1D Dyson gas by means of a path integral. For static eigenvalue configurations this was done  by Dean and Majumdar in~\cite{Dean:2006wk}. We generalize and extend their derivation to include the relaxation dynamics of DBM which adds a set of $N$ linear potentials to the Hamiltonian describing the evolving Dyson gas. Then, we perform an analytical saddle point evaluation of the path integral, which allows us to derive the time-dependent average eigenvalue distribution, given an initial fluctuated Hessian. `Time' here denotes the field displacement along the path in field space. Given this time-dependent eigenvalue distribution, we compute the saddle point action which gives us the transition probability as a function of distance in field space. 

These results are general for DBM in Gaussian random matrix theory, which itself has widespread applications beyond inflationary cosmology, including in recent years in areas like image analysis, genomics, epidemiology, engineering, economics and finance, for reviews see e.g.~\cite{ANU:298726,Bouchard:2009qf}. Then, we specify our results to an ensemble of Hessians with an eigenvalue distribution describing inflationary critical points (the lightest mass eigenvalues are very slightly tachyonic to describe slow-roll). Computing the probability of a graceful exit from such a random inflationary critical, we find our central result that the exit probability for small-field inflation in the landscape is exponentially small. The suppression exponent increases quadratically with number of light fields $N$.

We then compare this behaviour of small-field inflation in the landscape with large-field models, whose underlying structure and/or symmetry usually guarantees the existence of viable post-inflationary minimum. Taken at face value, this implies a strong exponential bias against small-field inflation being the dominant regime in the landscape. 

Finally, we discuss the influence of the $\exp(-c N^2)$ suppression of small-field inflation on the probability of observing negative spatial curvature in a landscape where the various dS vacua and inflationary critical points are populated via Coleman-De Luccia (CDL) tunneling transitions. Following the methodology of~\cite{Freivogel:2005vv}, the exponentially strong dependence on the number $N$ of light fields participating in a small-field inflationary critical point leads exponentially strong posterior probability distribution function for $N$ derived from the non-observation of spatial negative curvature. This severely limits the effective number such light fields to $N\ll 10$.


\section{The static ensemble}\label{sec:static}

In this section we introduce the basic concepts from Random Matrix Theory which we will apply to landscape statistics. The fundamental premise is that the scalar's mass matrix belongs to a classic ensemble in RMT, for which limiting eigenvalue distributions, fluctuation probabilities and other useful properties  are known.  We review how RMT tools allow us to determine the ratio between minima and flat inflection points in a  toy-model landscape.


\subsection{Basic concepts}
In what follows we assume that the mass matrix in the string landscape belongs to a Gaussian ensemble. Though this approximation is very restrictive and fails to incorporate some of the rich structure of the string landscape   \cite{Denef:2004ze,Denef:2004cf,Marsh:2011aa,Rummel:2013yta,Long:2014fba,Brodie:2015kza,Sousa:2014qza,Marsh:2015zoa}, we argue that it still retains some fundamental features that allow for a qualitative understanding of the landscape's properties while at the same time permitting the use of the very developed RMT machinery for this class of ensembles \cite{Aazami:2005jf,Easther:2005zr,Pedro:2013nda}.

The Gaussian ensembles are defined as sets of orthogonal, unitary or symplectic matrices whose entries are independent and identically distributed (i.e.random) variables drawn from some distribution $\Omega(\mu,\sigma)$. The observed universality property of RMT implies that the properties of the ensemble of matrices $M$ are insensitive to $\Omega$, provided its moments are appropriately bounded.

The probability of observing a given matrix $M$ with eigenvalues $\{\lambda_i\}$ in a Gaussian ensemble can be found by integrating the probability density function $dP$ 
\be
dP=\mathcal{C} \ \exp \left(-\frac{\beta}{2 \sigma^2}\sum_{i=1}^N \lambda_i^2 +\beta \sum_{i<j}^N\ln \left|\lambda_i -\lambda_j\right |\right) \prod_{i=1}^{N} d\lambda_i
\label{eq:pdfStatic}
\ee
over the eigenvalues of $M$, such that
\be
P(M)=\int dP,
\ee
where $\mathcal{C}$ is a $N$-dependent constant that can be found by performing unconstrained integration of the probability density function \eqref{eq:pdfStatic} and demanding
\be
\int_{-\infty}^{+\infty}dP\equiv 1
\ee
A more in depth derivation of Eq. \eqref{eq:pdfStatic} and associated concepts can be found in \cite{mehta}.

Equation \eqref{eq:pdfStatic} is identical to the partition function of a one dimensional gas of charged particles  executing Brownian motion under the influence of a quadratic self-interaction and a repulsive logarithmic potential, a fact that was first noted by Dyson \cite{Dyson:1962es} and that allows for an intuitive understanding of the behaviour of the coupled system of eigenvalues. The equilibrium eigenvalue density function, which gives the probability of finding an eigenvalue in the interval $[\lambda, \lambda+d \lambda]$, can be derived from Eq. \eqref{eq:pdfStatic} and is the well know Wigner semi-circle
\be
\rho(\lambda)=\frac{1}{\pi N \sigma^2}\sqrt{2 N \sigma^2-\lambda^2},
\label{eq:WignerSC}
\ee
which has support in the interval $\lambda\in[-\sigma \sqrt{2N},\sigma \sqrt{2N}]$. In what follows we set $\sigma^2=2/N$ in order to have a clear definition of the physically relevant spectra that is independent of the dimensionality of the field space. This choice implies that the masses in the equilibrium spectrum are distributed in the interval $[-2,2]\ M_P$. Formally Eq. \eqref{eq:WignerSC} can be obtained by finding the eigenvalue configuration that maximises Eq. (\ref{eq:pdfStatic}) in the limit of large $N$, and is a particular case of the computation presented in the next section.

\begin{figure}[h]
\begin{center}
\includegraphics[width=0.6\textwidth]{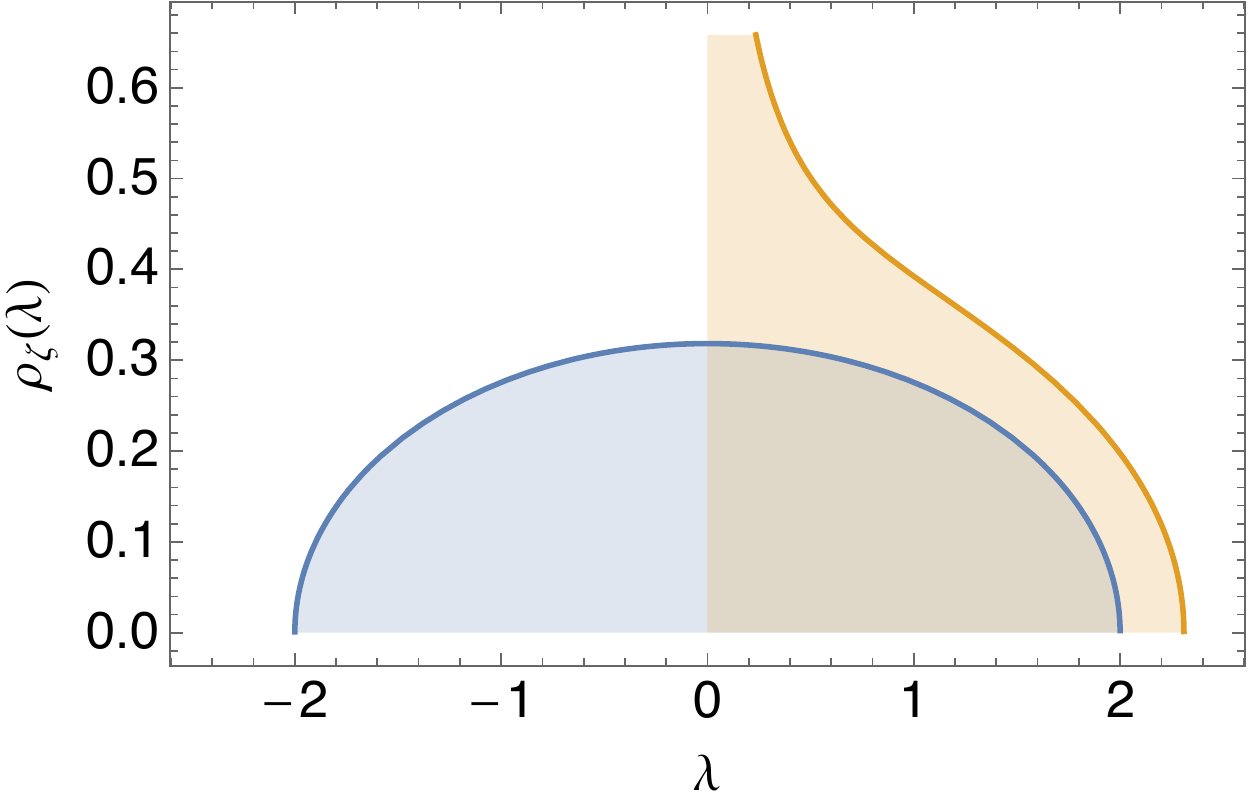}
\caption{Gaussian ensemble spectra: in blue: typical spectrum, following the semi-circle law of Eq. \eqref{eq:WignerSC}. Yellow: fluctuated spectrum with all eigenvalues positive, following the distribution law of Eq. \eqref{eq:distLaw} with $\zeta=0$.}
\label{fig:1}
\end{center}
\end{figure}

The semi-circle distribution, centred around the origin implies that a typical point in the Wigner landscape has on average half of the directions tachyonic, which is unsuitable to describe local minima and inflationary inflection points in the landscape. Such anthropically relevant points in the landscape  correspond therefore to fluctuations away from the equilibrium configuration, prompting the question of how frequently do these rare mass matrices arise and what is their eigenvalue spectrum. Both these questions have been addressed in the RMT literature, in particular Dean and Majumdar showed in \cite{Dean:2006wk} that spectra with all eigenvalues larger than $\zeta$ follow the large N distribution law
\be
\rho_\zeta(\lambda)=\frac{1}{6 \sqrt{3} \pi }\left(3 \lambda-\zeta +\sqrt{12+\zeta ^2}\right) \sqrt{\frac{-3 \lambda+\zeta +2 \sqrt{12+\zeta ^2}}{\lambda-\zeta }},
\label{eq:distLaw}
\ee
which asymptotes to Eq. \eqref{eq:WignerSC} in the limit $\zeta\rightarrow-2$, c.f. Fig. \ref{fig:1}. \\

\begin{figure}[h]
\begin{center}
\includegraphics[width=0.6\textwidth]{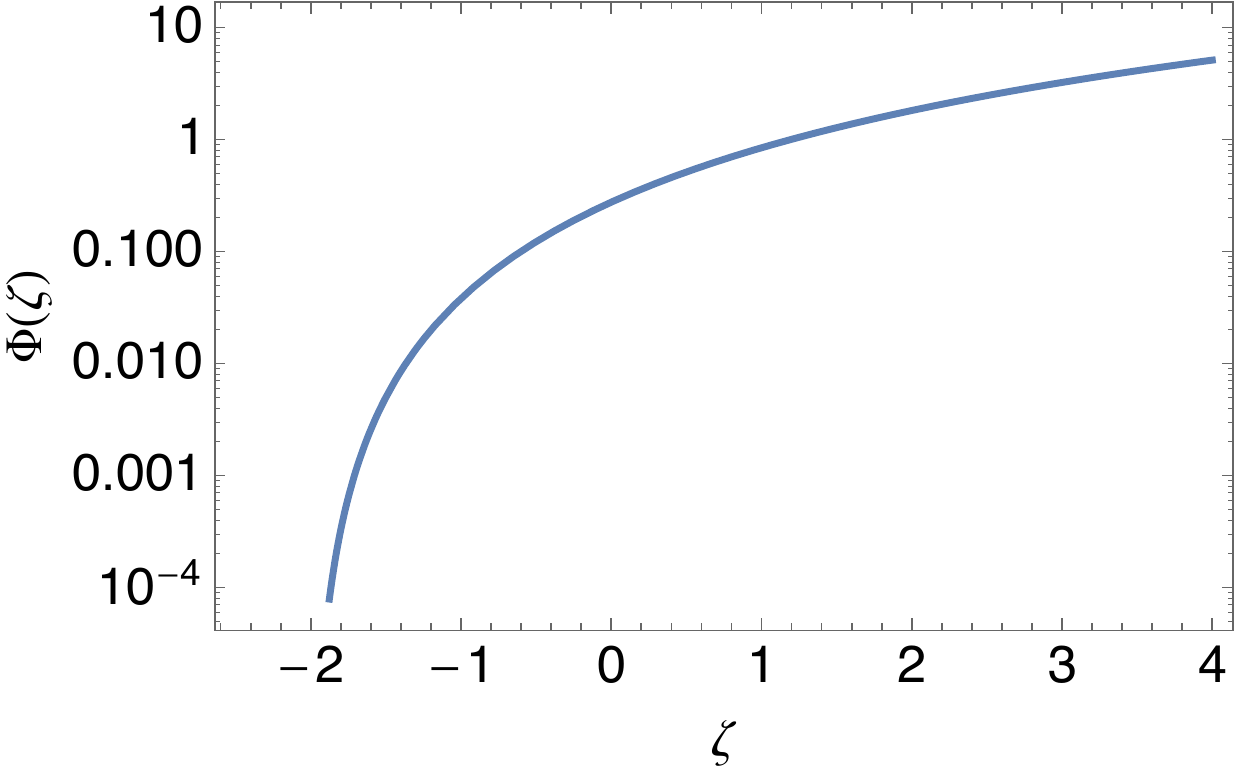}
\caption{Dean and Majumdar's rate function $\Phi(\zeta)$ defined in Eq. \eqref{eq:DM}. The rapid growth of $\Phi$ away from $\zeta=-2$ implies an exponential suppression of the probability of occurrence of such spectra in Gaussian ensembles.}
\label{fig:rate}
\end{center}
\end{figure}

The probability of such  fluctuations can be computed by integration of the probability density function
\be
P(  \lambda_i>\zeta\ , \forall\ i)=\int_{\zeta}^{+\infty}dP,
\ee
whose saddle point evaluation led the authors of  \cite{Dean:2006wk} to the result
\be
P(  \lambda_i>\zeta\ , \forall \ i)= \exp\left(-\beta N^2 \Phi(\zeta)\right),
\ee
where the rate function is given by
\be
\Phi(\zeta)=\frac{1}{432} \left[72 \zeta ^2-\zeta ^4+\left(30 \zeta +\zeta ^3 \right)\sqrt{12+\zeta ^2}+108 \ln36-216 \ln\left(-\zeta +\sqrt{12+\zeta ^2}\right)\right].
\label{eq:DM}
\ee
We plot $\Phi(\zeta)$ in Fig.\ref{fig:rate}. Note the rapid growth of the rate function away from the edge of the equilibrium distribution ($\zeta=-2$), implying that spectra with all eigenvalues significantly greater than $-2$ are very rare events in the Wigner landscape. One therefore concludes that both minima and flat inflection points do not abound.\\

\subsection{Consequences for landscape statistics}\label{sec:staticland}

The previous results can be used to assess the relative abundance of minima and inflationary flat points in the Wigner landscape. This was the focus of \cite{Pedro:2013nda}, the results of which we now review.

Let us start by defining the physically relevant spectra: minima and inflationary points. Minima are defined as points in field space where all eigenvalues are positive and larger than a reference value $\eta$. They occur with a probability given by 
\be
P(\mathrm{inf})=P(\lambda_i>\eta\ , \forall \ i)=e^{-\beta N^2 \Phi(\eta)}.
\ee
Inflationary spectra are taken to be those with at least one eigenvalue in the interval $[-\eta,\eta]$ and with all the remaining masses above $\eta$. The probability for such spectra is then given in terms of the rate function as 
\be
P(\mathrm{inf})=P(\lambda_i>-\eta \ , \forall \ i)-P(\lambda_i>\eta\ , \forall\ i)=e^{- \beta N^2 \Phi(-\eta)}-e^{- \beta N^2 \Phi (\eta)}.
\label{eq:Pinf}
\ee
From these we see that inflationary inflection points are more abundant than minima in the Wigner landscape by a factor of 
\be
\#_{av.-min}^{saddle}(inf)\equiv\frac{P(\mathrm{inf})}{P(\mathrm{min})}=e^{-\beta N^2 \{\Phi(-\eta)-\Phi(\eta)\}}-1\quad.
\ee
Here we define $\#_{av.-min}(inf)$ to be the number of inflationary saddle points per minimum, averaged over all minima. This expression can be approximated by expanding the rate function for small $\eta$ as
\be
\#_{av.-min}^{saddle}(inf)\sim e^{\beta N^2 \eta\Delta\: },\qquad\text{where}\qquad \Delta=2\left.\frac{d\Phi}{d\eta}\right|_{\eta=0}=\frac{4}{3 \sqrt{3}} +\mathcal{O}(\eta).
\ee
These considerations show that in this simple landscape there are exponentially more field space regions where inflation can occur than local minima where it can end. 

The formalism employed in this discussion allows us to count relative abundance of certain mass spectra, however it is unable to tell us the likelihood of connecting an inflationary trajectory with a post-inflationary minimum, for that one needs to go beyond the static ensemble and consider its 'time' dependent generalisation: Dyson Brownian motion. We perform the analytical computation of DBM in the next section, generalising the method of \cite{Dean:2006wk,Dean:2008} to the time dependent Coulomb gas. The reader interested only in the physical implications may safely skip to section \ref{sec:exit} where we use the results of section \ref{sec:DBM} to determine the exit probability in a Wigner landscape and discuss the consequences of our results.


\section{Dyson Brownian motion: the computation}\label{sec:DBM}

In the formalism of the previous section different eigenvalue configurations correspond to distinct equilibrium states for the Coulomb gas. There is no notion of time evolution or dynamics, the velocities of the point charges play no role in the physics of the gas. In his seminal work \cite{Dyson:1962} Dyson  introduced ``time'' evolution in this context by postulating that the point particles/eigenvalues were undergoing stochastic Brownian motion (rather than deterministic Newtonian evolution). In Dyson's approach ``time'' is not necessarily related to physical time, in fact for the applications we are interested in it is to be identified with field space distance \cite{Marsh:2013qca}. In this picture, which became known as Dyson Brownian motion, the entries of a matrix $M$, belonging to one of the Gaussian ensembles defined earlier, evolve in ``time'', $s$, according to $M_{ij}(s+\delta s)=M_{ij}(s)+\delta M_{ij}$,  where fluctuations display the following statistical properties
\bea
\langle\delta M_{ij}\rangle=-M_{ij}\frac{\delta s}{\sigma^2 f}\label{eq:drift}\ ,\label{eq:DBM1}\\
\langle(\delta M_{ij})^2\rangle= (1+\delta_{ij}) \frac{\delta s}{\beta f}\label{eq:diff}\ .
\label{eq:DBM2}
\eea
This implies that the $M_{ij}$ are undergoing simple uncoupled Brownian motion, interacting only with themselves and the medium in which they move. Note that Eq. \eqref{eq:drift} parametrises the drift and Eq. \eqref{eq:diff} the diffusion of the matrix Brownian motion. It is useful to note that Eqs. \eqref{eq:drift} and \eqref{eq:diff} allow us to write the infinitesimal shift in each matrix component as 
\be
\delta M_{ij}=A_{ij}-M_{ij}(s) \frac{\delta s}{\sigma^2 f}\ ,
\label{eq:matrixDBM}
\ee
where $A_{ij}$ are zero mean stochastic variables, describing the interaction between the matrix elements $M_{ij}$ and the medium in which they propagate and the last term is a restoring force. Equation \eqref{eq:matrixDBM} will be useful in the implementation of the numerical evolution of the matrix DBM performed in Sec. \ref{sec:exit} as a means of checking the analytic results we are about to derive.  The Brownian motion of matrix elements $M_{ij}$ was shown to be in one-to-one correspondence with the Brownian motion of the eigenvalues $\lambda_i$ of $M$ \cite{Dyson:1962}.

The time evolution described by Eqs. \eqref{eq:drift} and \eqref{eq:diff} can equivalently be formulated in terms of a probability density function $d P(M_{ij},s)$ \footnote{A note on notation: In contrast to the classical references \cite{Dyson:1962, mehta, Uhlenbeck:1930zz} where the probability density function is denoted by $P$, we choose to denote it by $d P$, reserving $P$ for the actual probability, obtained by integration of the density function $d P$.  } which is the solution to the Smoluchowski equation \cite{Uhlenbeck:1930zz}
\be
f \frac{\partial\ d P}{\partial s} =\sum_{i,j=1}^N\left\{\frac{\beta}{2}(1+\delta_{ij}) \frac{\partial^2\ dP}{{\partial M_{ij}}^2}+\frac{1}{\sigma^2}\frac{\partial}{\partial M_{ij}}(M_{ij} dP)\right\}\ .
\label{eq:Smolu}
\ee
The transition probability between an initial mass matrix $M_0$ and a final configuration $M$, over a distance $s$ in field space is therefore given by
\be
P(M(s),M_0)=\int d P\ ,
\ee
where the joint probability density function is the solution to Eq. \eqref{eq:Smolu}  \cite{Uhlenbeck:1930zz}
\be
d P=\mathcal{C} \ \exp \left\{-\frac{\beta}{2 \sigma^2(1-q^2)}\mathrm{Tr}[(M-q M_0)^2] \right\} dM_{ij}\ ,
\label{eq:jointpdf}
\ee
with $q=\exp(-s/(\sigma^2 f))$ being the field space distance measured in units of the correlation length $\Lambda_h\equiv\sigma^2 f$. For the moment we set $\sigma^2=a/N$ but will ultimately choose to normalise the mass spectra as in Sec. \ref{sec:static} by setting $a=2$. The overall factor $\mathcal{C}$ is a $q$-dependent function ensuring the correct normalisation of the fluctuation probabilities:
\be
\int_{-\infty}^{+\infty} d P \equiv 1.
\ee

Computation of transition probabilities between a given pair $(M_0,M(s)) $ can therefore be obtained via integration of Eq. \eqref{eq:jointpdf}. While numerical methods can be readily applied (though not without inherent limitations like the need to work at relatively small $N$), one can also use analytical methods in the large N limit and approximate the integral by its saddle point.


We start by analysing Eq. \eqref{eq:jointpdf} from the perspective of the ''time'' dependent Coulomb gas. Expanding the trace in the joint pdf one finds
\be
\begin{split}
\text{Tr}[(M-q M_0)^2] =&\mathrm{Tr}[M^2]-2 q\ \mathrm{Tr}[M M_0]+q^2\ \mathrm{Tr}[M_0^2] \\
=&\sum_{i=1}^{N} (\lambda_i^M)^2 -2 q \lambda_i^M M_0^{i i}+q^2  (\lambda_i^{M_0})^2\ ,
\end{split}
\ee
where $M_0^{ii} $ denotes the $i$-th diagonal element of the initial matrix $M_0$. Note that the last term on the r.h.s. is independent of the integration measure and therefore gives rise to an irrelevant overall constant in the partition function, which we will ignore in the ensuing discussion. One then sees that in the time-dependent Coulomb gas picture, the charged particles are subject to a quadratic self interaction  and to a linear potential. Note that in the static ensemble only the quadratic term was present and also that in the DBM context both are ''time'' dependent. Furthermore the memory of the initial conditions enters only through the strength of the linear potential and decreases with ``time'' or equivalently as we move away from the initial field space position.

The Hamiltonian for the ''time'' dependent Coulomb gas is then given by 
\be
H(\lambda)=-\frac{\beta}{2}\sum_{i=1}^{N} \left\{ \frac{\lambda_i^2}{\sigma^2 (1-q^2)}-\frac{2 q \lambda_i M_0^{ii}}{\sigma^2 (1-q^2)}\right\}+\frac{\beta}{2}\sum_{i\neq j} \ln|\lambda_i-\lambda_j |\ ,
\label{eq:Hlambda}
\ee
where the logarithmic interaction arises from the Jabobian of the coordinate change $d M_{ij}\rightarrow d \lambda_i$ \cite{mehta}.  In Eq. \eqref{eq:Hlambda} each eigenvalue is subject to a different linear potential, whose strength is set by the diagonal entries of the initial state matrix $M_0$. Given that the eigenbasis of $M$ and $M_0$ are in general not alligned, the diagonal entries of $M_0$ do not correspond to its eigenvalues, but instead one has
\be
M_0=\mathcal{U}\ \mathrm{diag}(\lambda_{M_0}^1,... , \lambda_{M_0}^N)\ \mathcal{U}^\dagger\ ,
\label{eq:M0}
\ee
where $\mathcal{U}$ is the matrix that diagonalises $M_0$, such that
\be
M_0^{ii}=\sum_{j=1}^N \mathcal{U}^{ij}\mathcal{U}^{\dagger ji} \lambda_{M_0}^{j}\ .
\ee
In order to proceed one must assume that all eigenvalues are subject to the same linear force. This corresponds to setting  $M_0^{ii} = m$, i.e., approximate the $N$ diagonal entries of the initial matrix by a single variable $m$. A natural choice is to take $m$ to be the average of the $M_0^{ii}$. In doing so one sees that the average force is given by the mean eigenvalue of $M_0$ since
\be
m\equiv\frac{1}{N}\sum_{i=1}^N M_0^{ii}=\frac{1}{N} \mathrm{Tr}[ M_0]\equiv\langle\lambda_{M_0}\rangle.
\label{eq:m}
\ee
One can therefore compute $m$ by taking the first moment of Eq. \eqref{eq:distLaw}
\be
m= \int d \lambda \ \lambda  \rho_{M_0}(\lambda)\ ,
\label{eq:m1}
\ee
where $\rho_{M_0}$ denotes the density function corresponding to the initial state matrix $M_0$. This simplification allows one to rewrite $H(\lambda)$ in terms of two variables only: $m$ and the empirical eigenvalue distribution
\be
\rho(\lambda)=\frac{1}{N}\sum_{i=1}^{N} \delta(\lambda-\lambda_i)\ ,
\ee
obeying $\int\rho(\lambda) d\lambda=1$. We then find that the Hamiltonian becomes a functional of $\rho$, satisfying
\be
 H[\rho]=-\beta N^2 \mathcal{E}[\rho]\ .
 \label{eq:H}
\ee
The ''energy'' functional in Eq. \eqref{eq:H} is defined by
\be
\mathcal{E}[\rho]=\frac{1}{2 \tilde{a}}\int\lambda^2\  \rho(\lambda)\ d\lambda+\frac{b}{2 \tilde{a}}\int\lambda\ \rho(\lambda)\ d\lambda-\frac{1}{2}\int \ln|\lambda-\tilde{\lambda}|\rho(\lambda)\rho(\tilde{\lambda})\ d\lambda\ d\tilde{\lambda} +\mathcal{O}\left(\frac{1}{N}\right).
\label{eq:E}
\ee
For the sake of shorter formulae we have defined $\tilde{a}\equiv a (1-q^2)$ and $b\equiv -2 q m$ and omitted the $\mathcal{O}\left(\frac{1}{N}\right)$ term coming from the subtraction of the self-energy in the logarithmic interactions in going from \eqref{eq:Hlambda} to \eqref{eq:H} \cite{Dean:2008}. Probabilities are now computed via the functional integral
\be
P(M(s),M_0)=\mathcal{C}\int  d[\rho] \ J[\rho] \exp\left\{ -\beta N^2 \mathcal{E}[\rho]\right\},
\label{eq:Prho}
\ee
where $J[\rho]$ denotes the Jacobian from changing from an integral over the eigenvalues $\lambda_i$ to a functional integral over density fields $\rho(\lambda)$ and implies \cite{Dean:2008}
\be
P(M(s),M_0)=\mathcal{C} \int  d[\rho] \delta\left( \int d\lambda \rho(\lambda)-1\right) \exp\left\{-N\int d\lambda\  \rho(\lambda) \ln\rho(\lambda) -\beta N^2 \mathcal{E}[\rho]\right\}.
\ee
Making use of a Lagrange multiplier $\alpha$ to enforce the correct normalisation of the eigenvalue distribution, one finally arrives to the ''energy'' functional 
\be
\Sigma[\rho]\equiv\mathcal{E}[\rho]+\alpha\left\{ \int d\lambda\ \rho(\lambda)-1\right\}\ ,
\ee
that must be minimised in order to have a large $N$ approximation of the transition probability. The problem is now to determine the eigenvalue density function $\rho_c(\lambda)$ corresponding to the constrained spectrum $\lambda_i >\zeta\ , \forall \ i$ that minimises $\Sigma[\rho]$, i.e.
\be
\frac{d \Sigma}{d \rho}\Big|_{\rho=\rho_c}=0 \qquad\Leftrightarrow\qquad
\frac{\lambda^2}{2 \tilde{a}}+\frac{b \lambda}{2 \tilde{a}}-\int_\zeta^\infty \ln|\lambda-\tilde{\lambda}|\ \rho_c(\tilde{\lambda})\ d\tilde{\lambda}=0\ .
\label{eq:rhoc}
\ee
Differentiating Eq. \eqref{eq:rhoc} with respect to $\lambda$ one finds
\be
\frac{\lambda}{\tilde{a}}+\frac{b}{2\tilde{a}}=\mathrm{P}\int_{\zeta}^\infty d\tilde{\lambda} \ \frac{\rho_c(\tilde{\lambda})}{\lambda-\tilde{\lambda}}\ ,
\label{eq:tricomi0}
\ee
where $\mathrm{P}$ stands for Cauchi's principal value of the otherwise ill defined integral. By performing a trivial shift of variables $x=\lambda-\zeta$, $\tilde{x}=\tilde{\lambda}-\zeta$,  Eq. \eqref{eq:tricomi0} can be recast in a form that allows for direct application of Tricomi's theorem \cite{tricomi} to determine the critical eigenvalue density function:
\be
\frac{x+\zeta}{\tilde{a}}+\frac{b}{2 \tilde{a}}=\mathrm{P} \int_0^L d\tilde{x}\  \frac{\rho_c(\tilde{x})}{x-\tilde{x}}\ .
\label{eq:tricomi1}
\ee
Note that in going from Eq. \eqref{eq:tricomi0} to Eq. \eqref{eq:tricomi1} we have assumed that $\rho_c(x)$ is only non-zero in the compact interval $[0,L]$, an assumption that must be checked a posteriori. As a consequence of Tricomi's theorem, one finds 
\be
\rho_c(x)=-\frac{1}{\pi^2 \sqrt{x(L-x)}}\left\{\mathrm{P} \int_0^L dx' \sqrt{x'(L-x')}\ \frac{x'+\zeta+b/2}{\tilde{a}}\ \frac{1}{x-x'} +C'\right\}\ ,
\ee
with the constant $C'$ being determined by requiring the vanishing of shifted eigenvalue density function at $L$ : $\rho_c(x=L)=0$:
\be
\rho_c(x)=\frac{1}{2 \pi \tilde{a}}\sqrt{\frac{L-x}{x}}[L+2(x+\zeta+b/2)]\ ,\qquad x\in [0,L]\ 
\label{eq:eigvaldensity}.
\ee
By imposing that $\rho_c(x)$ integrates to unity over the interval $[0, L]$,  one finds
\be
L=\frac{2}{3}\left[-(\zeta+b/2)+\sqrt{6\tilde{a}+(\zeta+b/2)^2}\right] \ .
\ee
The interpretation of $\rho_c(x)$ as the eigenvalue density function requires it to be positive definite everywhere. 
Imposing $\rho_c(x=0)>0$ one finds that consistency requires $\zeta>\zeta_{edge}$ with
\be
\zeta_{edge}\equiv -\sqrt{2 \tilde{a}}-\frac{b}{2} \ ,
\ee
that is the fluctuation probabilities computed here are applicable only when $\lambda_i >\zeta>\zeta_{edge}\ ,  \forall \ i$. For $ \lambda_i>\zeta<\zeta_{edge}\ , \forall \ i$, the probability is of order unity, since the system is free to evolve to its equilibrium configuration.

The minimal energy  of the eigenvalue system, corresponding to the configuration $\rho_c$, is therefore given by
\be
\Sigma[\rho_c]=\frac{\zeta^2+b\ \zeta}{4 \tilde{a}^2}+\frac{1}{4\tilde{a}}\int\lambda^2\ \rho_c(\lambda) \ d\lambda+\frac{b}{4 \tilde{a}} \int \rho_c(\lambda)\ \lambda\ d\lambda-\frac{1}{2}\int d\tilde{\lambda} \ln |\zeta-\tilde{\lambda}|\ \rho_c  (\tilde{\lambda})\ .
\ee
Performing the integrals one finds
\be\begin{split}
\Sigma[\rho_c]=\frac{1}{216 \tilde{a}^2}&\left\{ -2\tilde{\zeta}^4+(30 a \tilde{\zeta}+ 2 \tilde{\zeta}^3)\sqrt{6 \tilde{a}+ \tilde{\zeta}^2}+ 9 \tilde{a}(-3 b^2 +8 \tilde{\zeta}^2+9 \tilde{a})+\right . \\
&\left . + 27 \tilde{a}^2 \left[\ln1296-4 \ln(-\tilde{\zeta}+\sqrt{6 \tilde{a} +\tilde{\zeta}^2})\right] \right\}\ ,
\end{split}
\label{eq:Sigma}
\ee
where we have defined $\tilde{\zeta}\equiv\zeta+b/2$.

In order to compute the correctly normalised probability we note that the function $\mathcal{C}$ in Eq. \eqref{eq:jointpdf} admits the following saddle point approximation
\be
\mathcal{C}\simeq \exp\left\{-\beta N^2 \Sigma[-\infty]\right\}=\exp\left\{-\beta N^2 \Sigma[\zeta_{edge}]\right\},
\ee
where we abbreviated $\Sigma[\rho_c(\zeta)]=\Sigma[\zeta]$ and used $\rho_c(-\infty)=\rho_c(\zeta_{edge})\Rightarrow\Sigma(-\infty)=\Sigma(\zeta_{edge})$. The normalised transition probability then becomes 
\be
P(M(s),M_0)=\exp\left\{-\beta N^2(\Sigma[\zeta]-\Sigma[\zeta_{edge}])\right\} \ .
\ee
The rate function is defined as $\Psi(\zeta)\equiv \Sigma(\zeta)-\Sigma(\zeta_{edge})$, which using Eq. \eqref{eq:Sigma} can be shown to reduce to
\be\boxed{
\Psi(\tilde{\zeta})=\frac{1}{108  \tilde{a}^2}\left\{ 36 \tilde{a} \tilde{\zeta}^2-\tilde{\zeta}^4+(15 \tilde{a} \tilde{\zeta}+\tilde{\zeta}^3)\sqrt{6 \tilde{a}+\tilde{\zeta}^2}+ 27 \tilde{a}^2\left[\ln(72\tilde{a}) -2\ln(2(\sqrt{6 \tilde{a}-\tilde{\zeta}}-\tilde{\zeta}))\right]\right\}}\ ,
\label{eq:Psi}
\ee
where as before we have $\tilde{\zeta}\equiv\zeta+b/2$ and $b\equiv -2 q m$.

This is the main result of the present paper. It allows us to estimate the transition probability between a matrix $M_0$ characterised by the average size of its diagonal entries $m$ and a matrix $M(s)$ whose eigenvalues are larger than $\zeta$ as a function of the field space distance $s$:
\be
P(M(s),M_0)= \exp{\left[-\beta N^2 \Psi(\zeta)+\mathcal{O}(N)\right]}\ .
\label{eq:Pfinal}
\ee
As a sanity check, note that in the limit of infinite separation in field space $s\gg \Lambda_h \Leftrightarrow q\rightarrow 0$, we recover the rate function of the static ensemble, Eq. \eqref{eq:DM}, as $\Psi(\zeta)\rightarrow \Phi(\zeta)$.


\subsection{Examples}

Let us now test the validity of Eqs. \eqref{eq:Psi} and \eqref{eq:Pfinal} by comparing their estimates for the transition probabilities for the orthogonal, unitary and symplectic ensembles, with numerical integration results. We assume two different initial states: one with $ \lambda_i>1\ , \forall\ i$, and another with $\lambda_i>-1\ , \forall\  i$, and in both cases look for fluctuations towards spectra with $\lambda>0\ , \forall \ i$. One can use the one dimensional gas interpretation of the system to get some intuition as to how it should behave. For the first type of transition, we expect the probability to be of order unity within the first correlation length, since this corresponds to relaxation of the system towards its most probable configuration. For fluctuations of the second type we expect the probability to drop rapidly at large $q$ as these transitions go against the natural flow of the system. In both regimes as $q$ decreases and memory of the initial conditions becomes fainter, the probability should drop, approaching the result of Eq. \eqref{eq:DM} for the static ensemble at $q=0$. 

\begin{figure}[h!]
	\centering
	\begin{minipage}[b]{0.49\linewidth}
	\centering
	\includegraphics[width=0.95\textwidth]{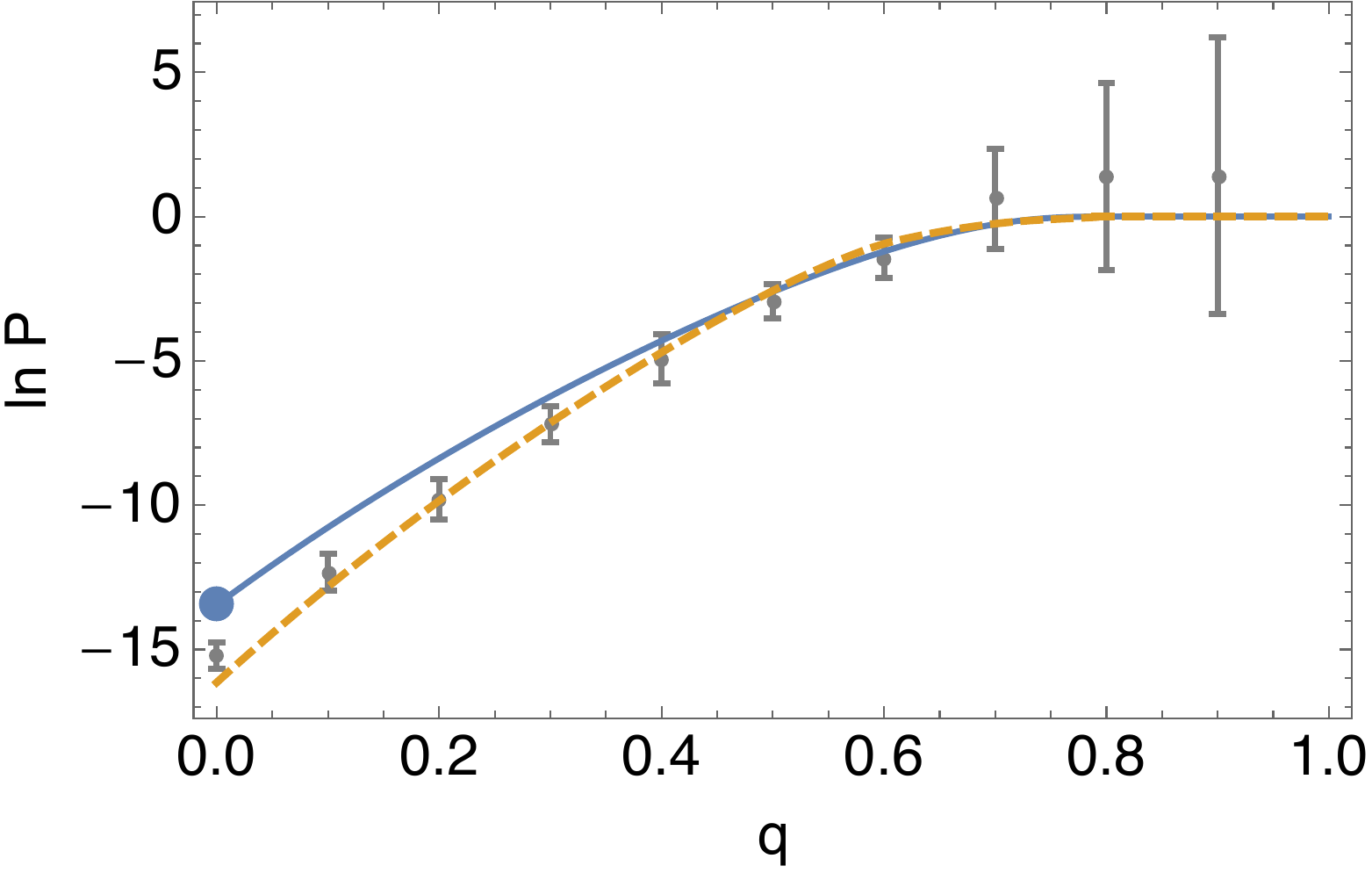}
    \end{minipage}
	\hspace{0.05cm}
	\begin{minipage}[b]{0.49\linewidth}
	\centering
	\includegraphics[width=0.95\textwidth]{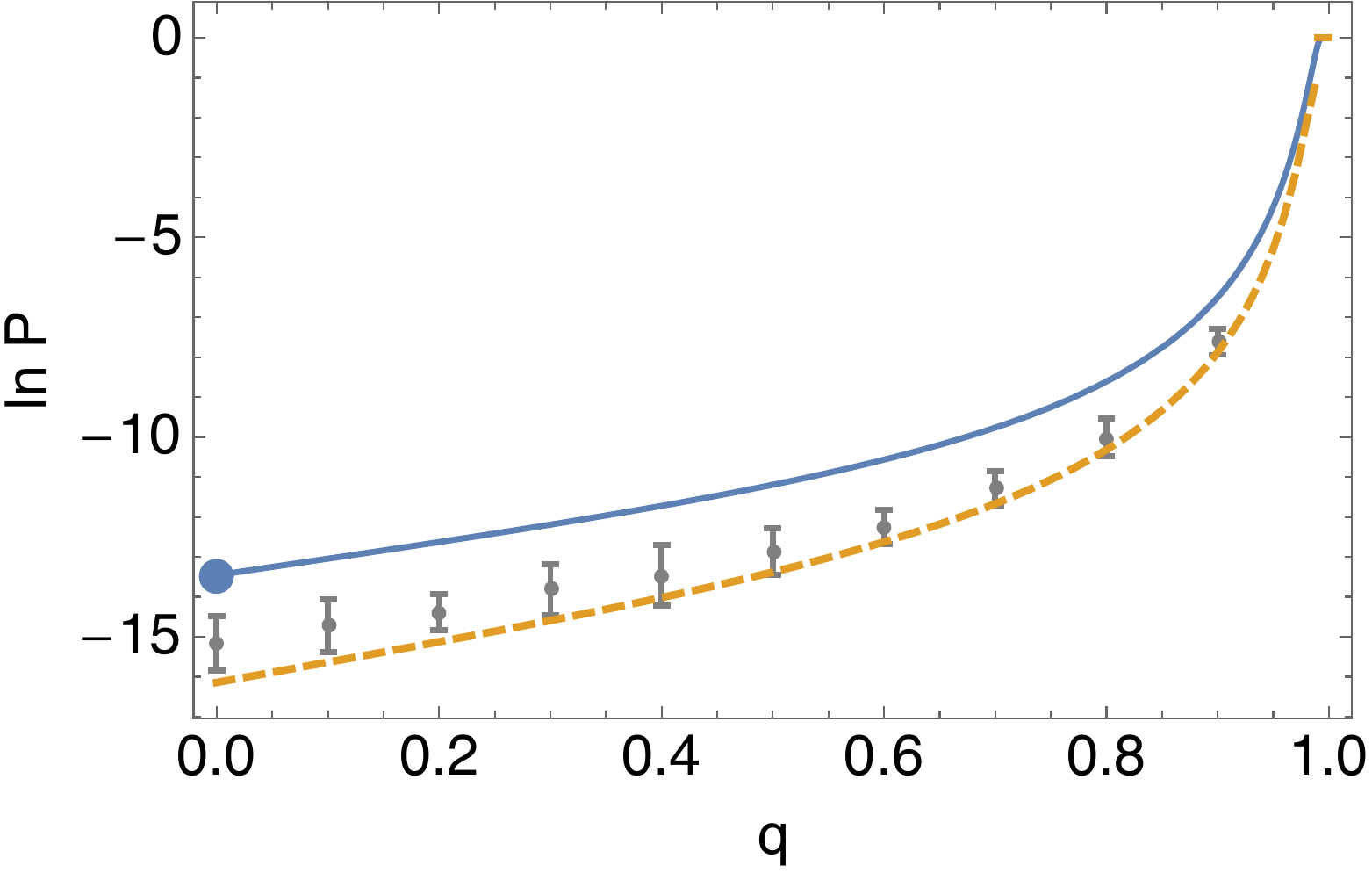}
	\end{minipage}
	\hspace{0.05cm}
\centering
	\begin{minipage}[b]{0.49\linewidth}
	\centering
	\includegraphics[width=0.95\textwidth]{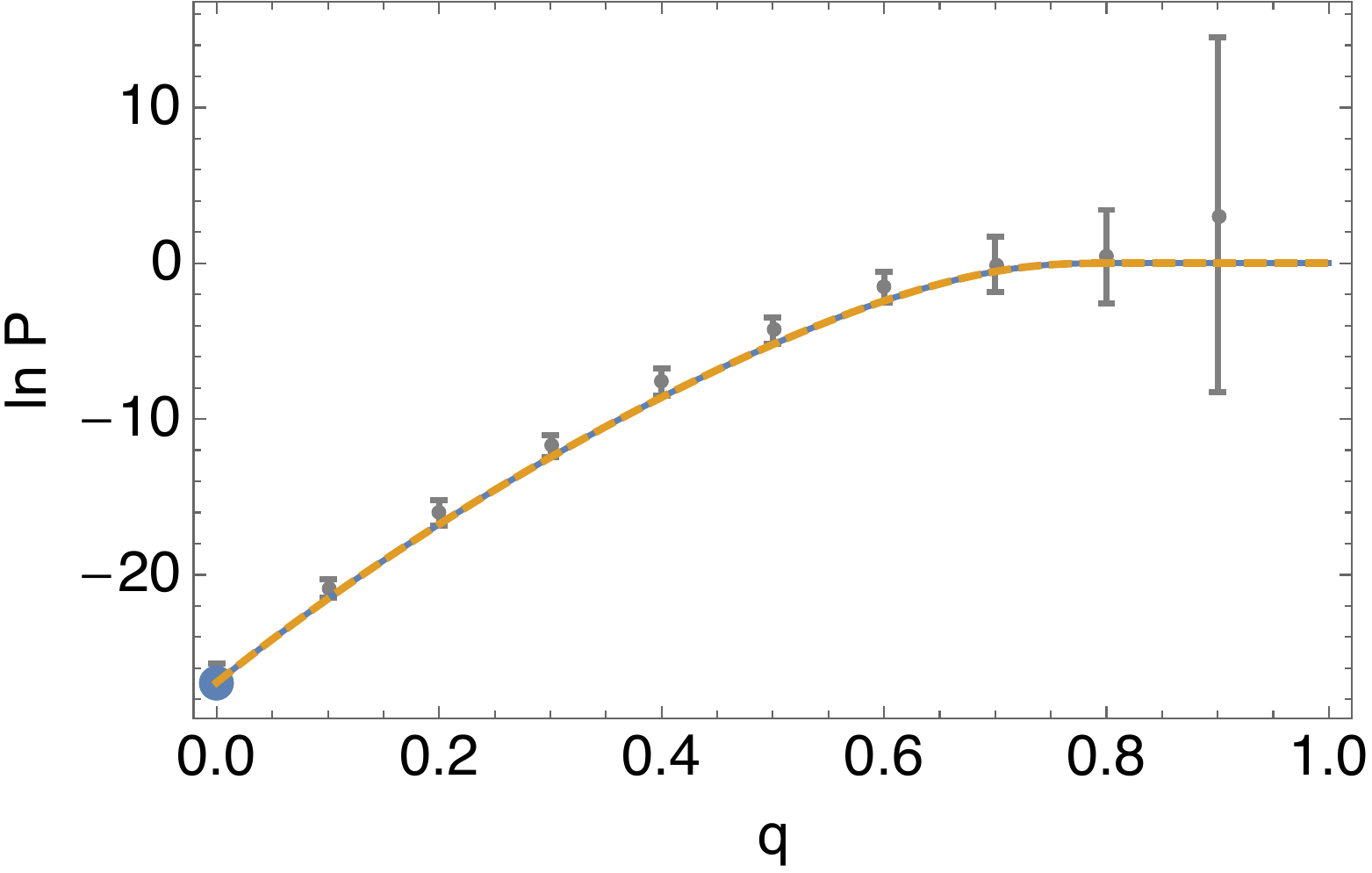}
    \end{minipage}
	\hspace{0.05cm}
	\begin{minipage}[b]{0.49\linewidth}
	\centering
	\includegraphics[width=0.95\textwidth]{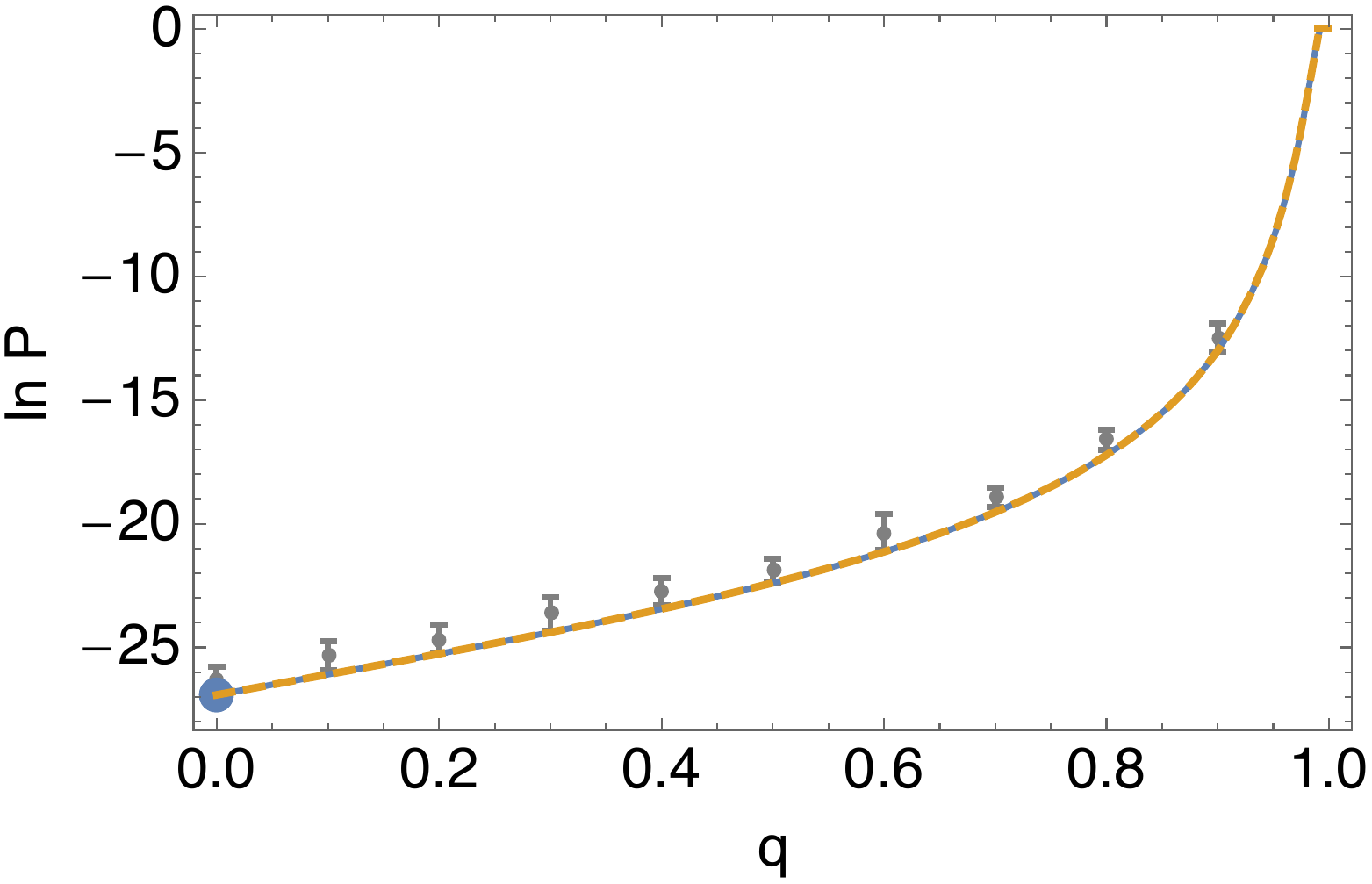}
	\end{minipage}
	\hspace{0.05cm}
	\centering
	\begin{minipage}[b]{0.49\linewidth}
	\centering
	\includegraphics[width=0.95\textwidth]{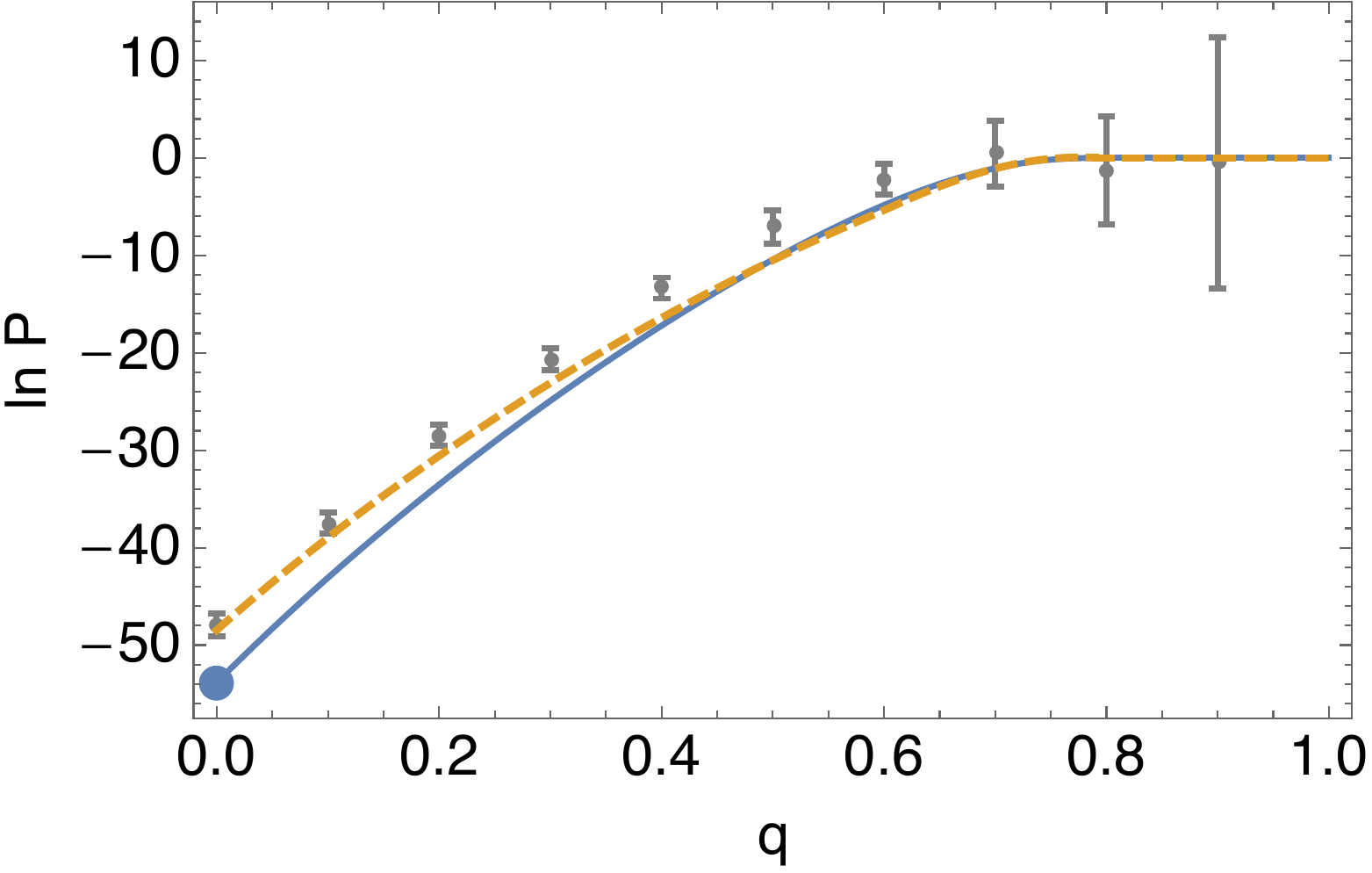}
    \end{minipage}
	\hspace{0.05cm}
	\begin{minipage}[b]{0.49\linewidth}
	\centering
	\includegraphics[width=0.95\textwidth]{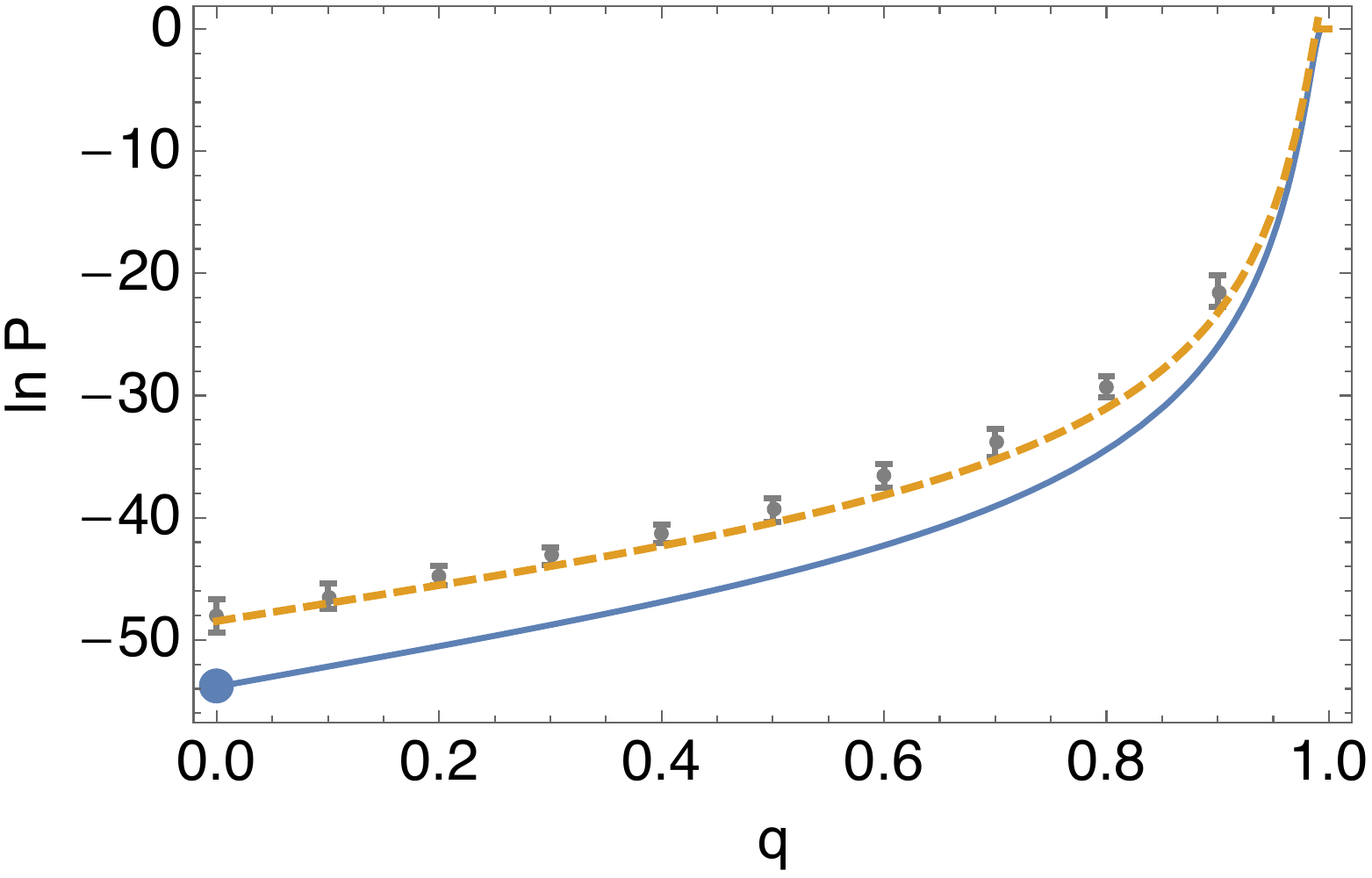}
	\end{minipage}
	\hspace{0.05cm}
\caption{Fluctuation probabilities for the orthogonal (top), unitary (middle) and symplectic (bottom) ensembles. The plots on the left correspond to fluctuation from spectra with $ \lambda_i>1 \ ,\forall \ i$ to spectra with $\lambda_i>0 \ ,\forall \ i$. The plots on the right correspond to fluctuation from spectra with $ \lambda_i>-1\ ,\forall \ i$ to spectra with $ \lambda_i>0\ ,\forall \ i$.  The points with error bars are obtained by numerical integration of the pdf, the blue/continuous line denotes the leading $N^2$ approximation, the yellow/dashed line includes the subleading $\mathcal{O}(N)$ corrections to the Hamiltonian. The blue circles represent the Dean and Majumdar's result for the static ensembles. We set $a=2$ for all cases.}
	\label{fig:tests}
\end{figure}

We compare the different estimates for the transition probabilities in Figure \ref{fig:tests} for $N=7$. Fluctuations from spectra with $\lambda_i>1\ , \forall \ i$, corresponding to a mean eigenvalue $m=1.52$ are presented in the left column, while the right column  gathers the results for initial states with $\lambda_i>-1\ , \forall \ i$, corresponding to $m=0.22$. We compare the numerical integration data with the saddle point approximation of Eqs. \eqref{eq:Psi} and \eqref{eq:Pfinal}. We see that our leading $\mathcal{O}(N^2)$ approximation captures the behaviour of the numerical integration for both fluctuations for all ensembles. Closer inspection reveals that Eqs. (\ref{eq:Psi}) and \eqref{eq:Pfinal} overestimate the transition probability for the orthogonal ensemble while they underestimate the probability for the symplectic ensemble. Keeping in mind that for large $N$ these discrepancies become irrelevant, we note nonetheless that they are due to the fact that in Eq. \eqref{eq:Psi} we neglected $\mathcal{O}(N)$ corrections of the form \cite{Dean:2008}
\be
\delta H=-N\left(1-\frac{\beta}{2}\right)\int d\lambda \  \rho(\lambda) \ln(\rho(\lambda)).
\ee
These terms originate from the subtraction of the divergent electrostatic self-interaction in going from Eq. \eqref{eq:Hlambda} to \eqref{eq:E} and from the Jacobian $J[\rho]$ in Eq. \eqref{eq:Prho}. Note that this $\mathcal{O}(N)$ correction vanishes exactly for the unitary ensemble ($\beta=2$), for which our $\mathcal{O}(N^2)$ approximation tracks the numerical integration data, c.f. Fig. \ref{fig:tests}. For the other ensembles, when one includes these subleading terms, assuming that they do not alter shape of the limiting eigenvalue density function $\rho_c$, one finds that the semi-analytic saddle point (dashed lines in Fig. \ref{fig:tests})  estimate falls within the error bars of the numerical integration data.\\

\begin{figure}[h!]
	\centering
	\begin{minipage}[b]{0.49\linewidth}
	\centering
	\includegraphics[width=\textwidth]{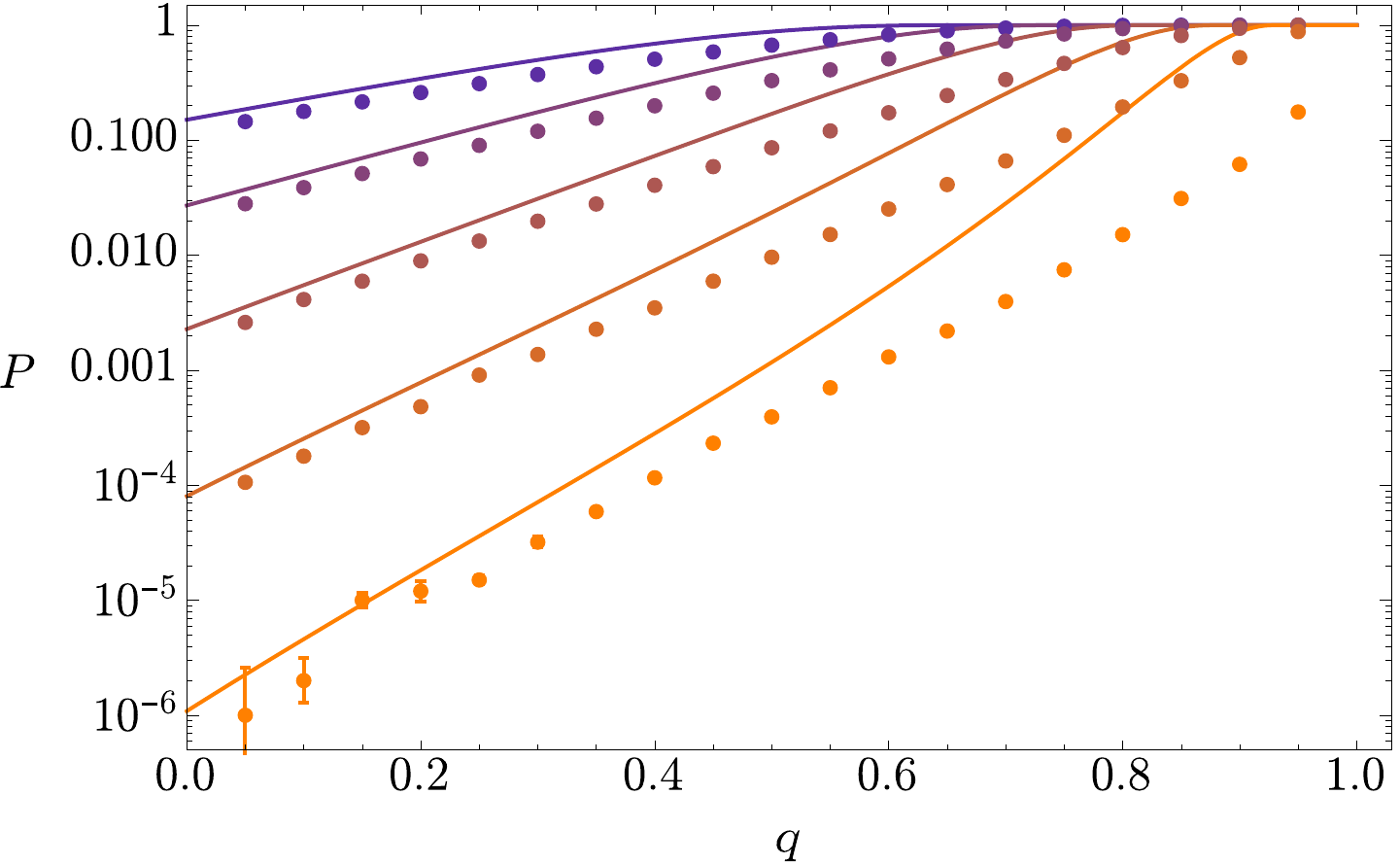}
    \end{minipage}
	\hspace{0.05cm}
	\begin{minipage}[b]{0.49\linewidth}
	\centering
	\includegraphics[width=0.95\textwidth]{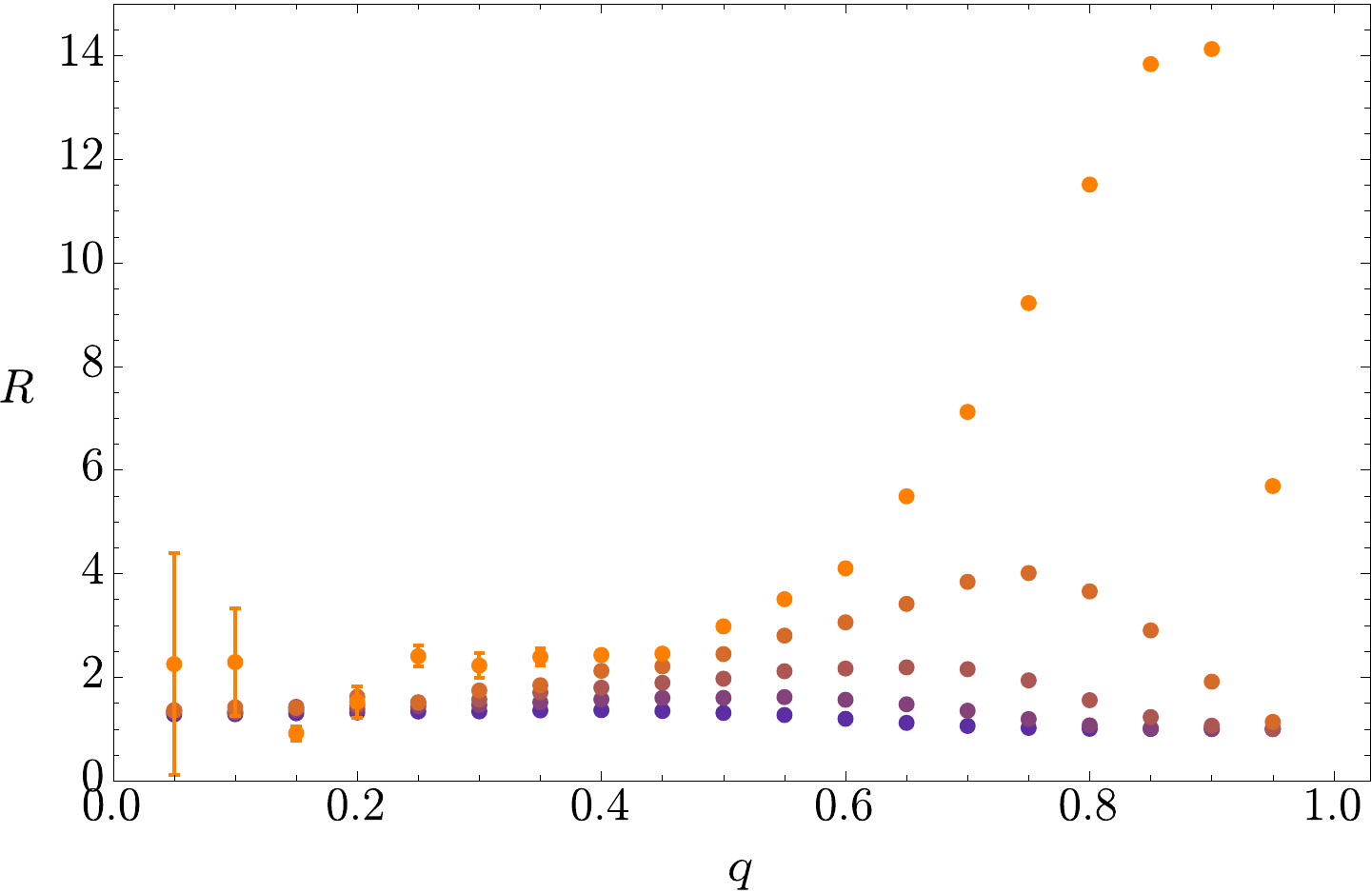}
	\end{minipage}
	\hspace{0.05cm}
\caption{{\it Left}: Transition probability as a function of "time" $q$, starting with a positive definite matrix and ending in various final states, see main text for more details. Dots correspond to the Brownian motion result whereas the lines represent the analytical estimate whose validity we are evaluating. {\it  Right}: Ratio between the analytic estimate of the transition probability and the Brownian motion result.}
	\label{fig:tests2}
\end{figure}

The results presented above clearly demonstrate  that the saddle point method is a good approximation to the integration of the PDF, once we make the the assumption of universality for the linear potential in Eq. \eqref{eq:Hlambda}. In order to fully understand the validity of our method for the computation of transition probabilities, one must test the robustness of this universality simplification. This can be easily done, for small $N$, by comparing the results of Eqs. \eqref{eq:Psi} and \eqref{eq:Pfinal} with actual random walks generated via Eqs. \eqref{eq:DBM1}-\eqref{eq:matrixDBM} for different initial and final conditions. In order to better assess the validity of our approximations we choose to work with the unitary ensemble $\beta=2$, since in this case the subleading $\mathcal{O}(N)$ corrections are absent and any deviation between the saddle point estimate and the numerical DBM has to be attributed to the limitations of the universal linear potential approximation. We choose initial spectra  $M_0 : \lambda_i > 0 \ , \forall \ i$ and look for fluctuations to spectra $M(s) :  \lambda_i > \zeta \ , \forall\ i$, with $\zeta =\{-1, -0.75, -0.5, -0.25, 0\}$. \footnote{There are two ways in which one can choose the $M_0$ that constitute the initial conditions for the Brownian motion: one can generate a large number of matrices of the desired ensemble, choose the ones that have the desired spectrum and perform the time evolution on each of those matrices. Alternatively one can compute the average eigenvalues of the matrices that have the right spectrum and  thus build a unique initial condition on which to run the Brownian motion. We choose the latter procedure and stress that the distinction between the two should  vanish in the large $N$ limit.  } Computational time constraints force us to work with $N=5$. The results are  presented in Fig. \ref{fig:tests2}, where one can see that the analytic estimate tracks the Brownian motion results for the various initial conditions, once enough "time" has passed. Typically after $1/2$ correlation length, $q<0.6$, the analytical estimate is within a factor of a few of the numerical result, as can be seen in the left panel of Fig. \ref{fig:tests2}.\\

\subsection{An analytic argument for a universal linear potential}
Beyond this clear numerical evidence, we will now provide analytical arguments justifying of the single linear-potential approximation and the universal $\Psi(q)=\Psi(0)-{\cal O}(q)\,,\,q\lesssim e^{-1}$ behaviour of the rate function. For this, we will use results of the saddle point computation, the time-dependent fluctuated eigenvalue density Eq.~\eqref{eq:eigvaldensity} and the time-dependent rate function Eq.~\eqref{eq:Psi}. 

We will perform our analysis perturbatively, by pushing the system away from its equilibrium configuration and then looking at the behaviour of the rate function for different slightly perturbed initial spectra, while selecting final states which  are small departures from the equilibrium configuration. In such a regime, spectra with eigenvalues larger than $\zeta_i=-2+\epsilon$ describe the initial conditions. Correspondingly we choose final spectra which have all eigenvalues larger than $\zeta_f=-2+\delta$. 

Next, we expand the rate function in small $\epsilon,\delta$ at small $q$, finding $\Psi=\frac{\delta^3}{12}\left(1+\frac32 q^2\right)-\frac{1}{16} \,\delta^2\epsilon^2 q $ to leading order. The constant piece reproduces the result of Dean and Majumdar for the static ensemble \cite{Dean:2006wk, Dean:2008}. Furthermore, we observe that the dominant $q$-dependent piece at small $q$ is linear in $q$. Its coefficient is uniquely determined by the edges of the fluctuated initial and final condition choices. 

Hence, for sufficiently small $q$ the rate function becomes linear in $q$. This explains the numerically observed universal behaviour at late times. Moreover, we recognize, that $\Psi$ becomes linear in $q$ typically within one correlation length. The exception to this case arises when we ask for the final conditions to be given by spectra more unlikely than the initial conditions, that is $\delta > \epsilon$. Once in such a regime, we see that the $q^2$ term in $\Psi$ will give way to the linear term only at successively smaller values of $q$. Comparing the analytic behaviour of $\Psi$ and the DBM result in Fig.~\ref{fig:tests2} around values of $q\simeq e^{-1}$ for the lowest set of curves where both initial and final conditions are given by all eigenvalues larger than $\zeta_i=\zeta_f=0$, we clearly observe the onset of the quadratic behaviour towards larger $q$.

We now proceed to consider the relaxation process of DBM in more detail. It is a basic feature of DBM that, whatever the initial conditions, relaxation drives the eigenvalue spectra to approach their static equilibrium configuration for $q\to 0$ at a rate estimated in \cite{Marsh:2013qca}. Now we conduct a gedanken experiment: assume that we stop the relaxation process momentarily at some small but otherwise arbitrary value $q=\tilde q\ll 1$. The  eigenvalue spectrum of a large set of matrices there is already close to the static equilibrium configuration. Thus, at $q=\tilde q$ we will find, with probability close to unity, only eigenvalue spectra of slightly fluctuated Wigner ensembles with a lower edge close to the semi-circle value $\zeta(\tilde q)=-2+\epsilon(\tilde q)$, where $\epsilon(\tilde q)$ depends on the spectrum at $q=1$. This slightly fluctuated Wigner ensemble thus forms the DBM-produced most probable \emph{local} initial conditions for relaxing towards further decreasing $q<\tilde q$. Now we unfreeze our system -- let the relaxation process resume and focus on the simplest case $N=2$ for extracting analytical results. The relaxation from $1$ to $\tilde q\ll 1$ has produced a spectrum of matrices $M_{\tilde q}$ which is given by a slightly fluctuated Wigner ensemble with edge at $\zeta(\tilde q)=-2+\epsilon(\tilde q)$. Upon unfreezing our system at $\tilde q$, this ensemble $M_{\tilde q}$ forms the $q=\tilde q$ initial conditions for further relaxation. Hence, these $q=\tilde q$ initial conditions given by the $M_{\tilde q}$-ensemble will relax further at $q<\tilde q$ driven by two linear potentials, and their average linear potential acting on $M_{\tilde q}$.

Next, we determine the two linear potentials and their variances acting on $M_{\tilde q}$ at $q=\tilde q\ll1$. They are given in terms of the spectrum of the $M_{\tilde q}$-ensemble describing the $q=\tilde q$ initial conditions. Hence, we compute $\langle\lambda_{M_{\tilde q}}^i\rangle\equiv \langle M_{\tilde q}^{ii}\rangle$ governing their strength, and their respective variances $\langle(M_{\tilde q}^{ii}-\langle\lambda_{M_{\tilde q}}^i\rangle)^2\rangle$. These quantities are the estimators of the two linear potentials and their variances, respectively. 

As we have seen in section \ref{sec:DBM}, the Hamiltonian contains $N$ linear potentials given by the $N$ diagonal entries $M_q^{ii}$. We want to determine the estimators above at $q=\tilde q\ll 1$. Hence, we can approximate the ensemble $M_{\tilde q}$ by the static fluctuated eigenvalue density $\rho(\mu, q=0$) with edge $\zeta(\tilde q)=-2+\epsilon(\tilde q)$. So, for $N=2$ we compute the estimators of two linear potentials by estimating $\langle M_q^{ii}\rangle$ from the static fluctuated eigenvalue density $\rho_c(\mu, q=0)$ of Eq.~\eqref{eq:eigvaldensity} as follows
\be\begin{split}
\langle \lambda_{M_q}^1\rangle &= \int \limits_{-2+\epsilon}^{\epsilon^2/8}\hspace{-1ex}d\mu\Big(\mu\,\rho_c(\mu,q=0)\Big)+{\cal O}(q)\\
\label{eq:Nflinearforces}
\langle \lambda_{M_q}^2\rangle &= \int \limits_{\epsilon^2/8}^{2+\epsilon^2/16}\hspace{-3ex}d\mu\Big(\mu\,\rho_c(\mu,q=0)\Big)+{\cal O}(q)\quad.
\end{split}
\ee
Here, matching to $M_{\tilde q}$ implies $\epsilon\equiv \epsilon(\tilde q)$. Then, the upper integration boundaries are determined to leading order in the lower edge shift $\epsilon$ by demanding equal probability for the left and right half-bands of the eigenvalue density  $\int_{-2+\epsilon}^{\epsilon^2/8}d\mu\rho_c(\mu,q=0)=\int_{\epsilon^2/8}^{2+\epsilon^2/16}d\mu\rho_c(\mu,q=0)=1/2+{\cal O}(\epsilon^3)$.

Using this prescription we get that $\langle \lambda_{M_0}^{1/2}\rangle=\pm  \frac{8}{3\pi}+{\cal O}(\epsilon^2)$ such that the total average eigenvalue is $\langle\lambda_{M_0}\rangle={\cal O}(\epsilon^2)$. Similarly, we can compute the variances of both $\langle\lambda_{M_0}\rangle$ and $\langle\lambda_{M_0}^i\rangle$. As in the case of the average eigenvalue, the variances deviate from their unfluctuated semi-circle values only at ${\cal O}(\epsilon^2)$. This behaviour is consistent with the earlier result that the ${\cal O}(q)$-term in the rate function describing the dominant linear potential term arises at ${\cal O}(\epsilon^2)$. 

Let us pause to review what we obtained. Our results show that the estimators for the eigenvalues giving the two linear potentials, $\langle\lambda_{M_0}^i\rangle$, as well as their variances deviate from the values for the unfluctuated semi-circle at ${\cal O}(\epsilon^2)$. This is an order \emph{lower} in $\epsilon$ than the deviation of the edge of the eigenvalue spectrum at small $q$ which is $\zeta(q)-\zeta(semi-circle)=\zeta(q)+2={\cal O}(\epsilon)$. Moreover, in the limit of the exact semicircle the mean of the two equal-size linear potentials vanishes. This implies that the effect of their equal-size variances of the semicircle distribution on the effective average linear potential must cancel out as well. Hence, the total values of the $N$ linear potentials and their variances do not determine the averaged linear potential. Instead, for small deviations from the semicircle it is the \emph{shift} of each of them driven by the ${\cal O}(\epsilon)$ shift of the spectral edge $\zeta$ which enters the averaged linear potential. 

We thus conclude that, at small $q$, \emph{the effects of having two linear potentials are given by just the average overall linear potential} given by $\langle\lambda_{M_0}\rangle=1/2 \sum_{i=1,2}\langle\lambda_{M_0}^i\rangle$ up to and including the second moments of the individual linear potentials at ${\cal O}(\epsilon^2)$. This result is important because it establishes that at small enough $q$ the single linear potential approximation becomes a good description, which a posteriori justifies the use of this approximation.


\section{Dyson Brownian motion: applications}\label{sec:Apps}

\subsection{The exit probability and exit-conditioned frequency count of random small-field inflation}\label{sec:exit}

Having solved the time dependent Coulomb gas in the previous section, we are now in a position to estimate the probability of connecting an inflationary point and a minimum in our simplified landscape. We adopt the definitions of Sec. \ref{sec:staticland} for minima and inflationary patches, namely
\be\begin{split}
\mathrm{minima}\qquad:&\qquad   \lambda_i>\eta \ ,\forall\ i; \\
\mathrm{inflation}\qquad:&\qquad  \lambda_i>-\eta\ \ ,\forall\ i \  \wedge \ \exists\  \lambda_i\in [-\eta,\eta]\ .
\end{split}
\label{eq:infmin}
\ee
Note that these definitions involve only properties of the mass matrix, and ignore the behaviour of the gradient and of the vev of the scalar potential, both of which play a role in the dynamics of the system. This simplification implies that our estimates for landscape transition probability are, strictly speaking, upper limits on that quantity: once we take the gradients and vevs into account, via the iterative process proposed in \cite{Marsh:2013qca}, the actual transition probability will surely be smaller than the values quoted here.

\begin{figure}[h]
\begin{center}
\includegraphics[width=0.65\textwidth]{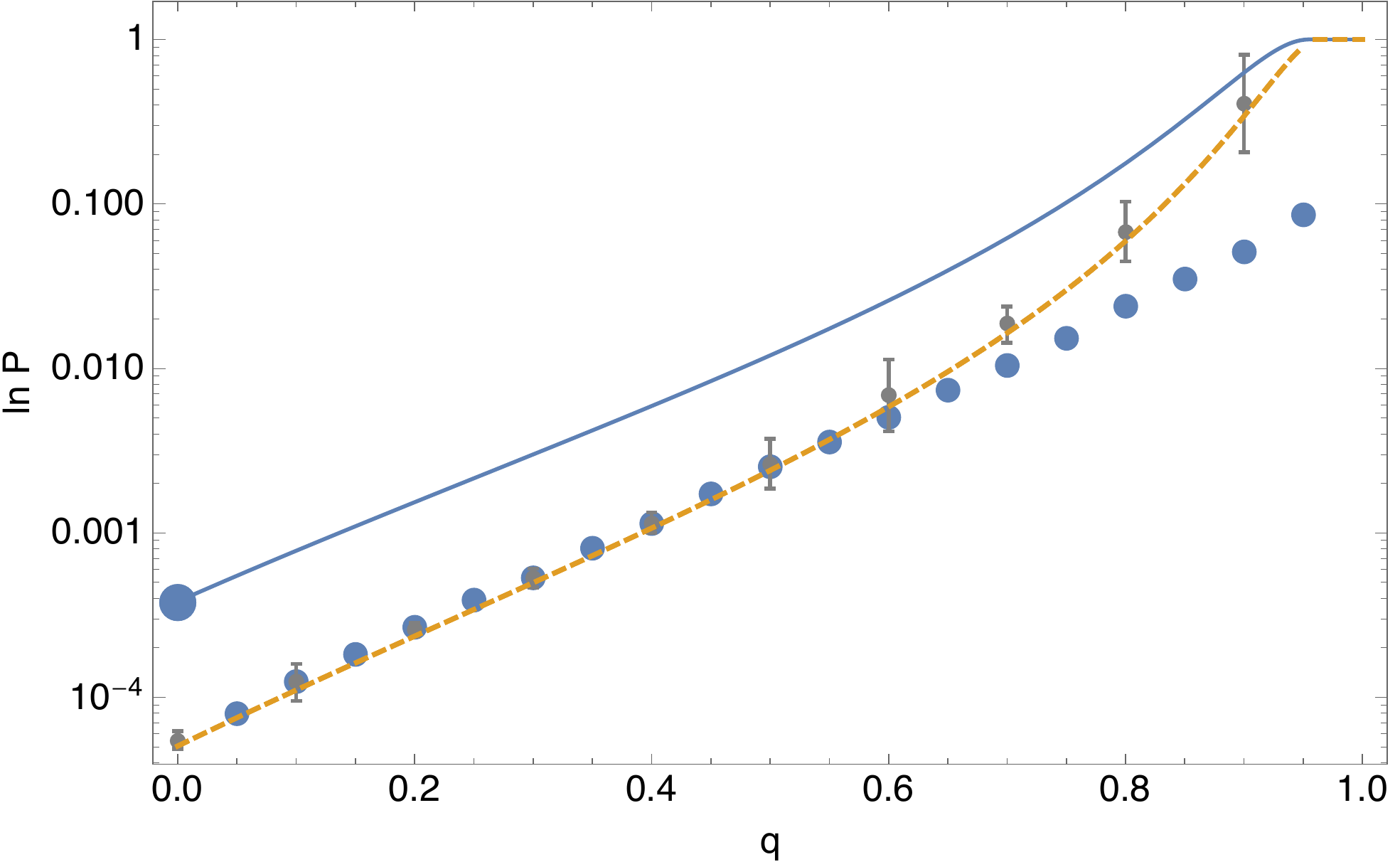}
\caption{Exit probability for the orthogonal ensemble ($\beta=1$) with $N=5$. Small circles denote the fluctuation probability estimated by performing a matrix random walk with $10^6$ matrices, the grey points with error bars correspond to numerical integration of the pdf, the blue/continuous line denotes the leading $N^2$ approximation, the yellow/dashed line includes the subleading $\mathcal{O}(N)$ corrections to the Hamiltonian. The large blue circle at $q=0$ represents the Dean and Majumdar's result for the static ensemble.}
\label{fig:Pexit}
\end{center}
\end{figure}

In what follows we use the same normalisation of the mass spectrum as in the static case, setting $a=2$ and for concreteness choose $\eta=0.1$. The definition of inflationary patches through Eq. \eqref{eq:infmin} implies that the mean linear potential acting on the eigenvalues is  $m=0.70$.  In Fig. \ref{fig:Pexit} we plot the exit probability for the orthogonal ensemble with $N=5$. \\

In order to have the correct picture of how likely a Universe that underwent a period of inflation ending in a minimum is, one has to take into account not only the transition probability discussed above but also the probability of having the correct starting point. The probability for the initial mass spectra, $P(\mathrm{inf})$, follows from the discussion of the static ensemble of Sec. \ref{sec:static}, in particular Eq. \eqref{eq:Pinf}. The likelihood of fluctuating to a minimum, having started with an inflationary spectrum can be estimated by recalling that the conditional probability of the two events is given by
\be
P(\mathrm{min}| \mathrm{inf})=\frac{P(\mathrm{min} \cap \mathrm{inf})}{P(\mathrm{inf})}.
\ee
Identifying the transition probability computed in the previous section with $P(\mathrm{min}| \mathrm{inf})$ one finds that the probability of having field space trajectories connecting inflationary patches with exit minima is given by
\be
P(\mathrm{min} \cap \mathrm{inf})=P(\mathrm{inf}) \ P(\mathrm{min}| \mathrm{inf}),
\ee
where $P(\mathrm{inf})$ is given by the static ensemble result of Eq. \eqref{eq:Pinf} and
 \be
 P(\mathrm{min}| \mathrm{inf})\simeq  \exp\left[- \beta N^2 \left\{ \frac{\ln 3}{4}+\frac{2 }{3 \sqrt{3}}( \eta- m q) \right\} \right]\quad,
 \ee
 where we performed an expansion of the time dependent rate function of Eq. \eqref{eq:Psi} to linear order in $\eta$ and $q$. Note that the linear approximation in $q$ is very accurate whenever  minimum and the inflationary patches are sufficiently separated in field space, that is when $q$ is sufficiently small, as can be seen in Fig. \ref{fig:Pexit}.\\
 
 \begin{figure}[t]
\begin{center}
\includegraphics[width=0.6\textwidth]{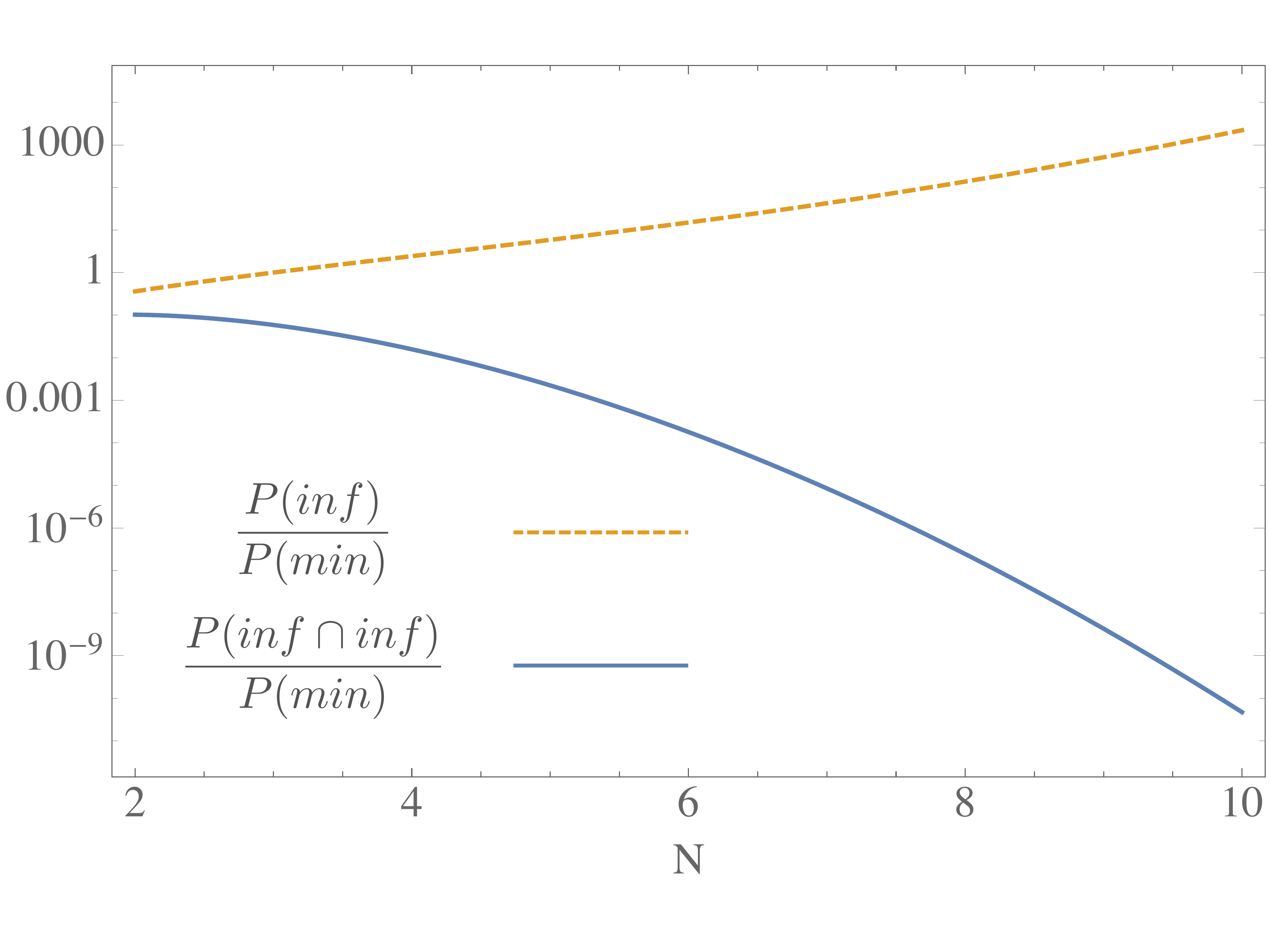}
\caption{The number of inflationary regions per minimum with (blue/continuous line) and without (yellow/dashed line) conditioning on a graceful exit, averaged over all minima.}
\label{fig:saddleCount}
\end{center}
\end{figure}
 
The probability of having field space trajectories connecting inflationary patches with exit minima is approximately expression given by 
\be
P(\mathrm{min} \cap \mathrm{inf})\simeq \exp \left[ -\beta N^2 \left(\frac{\ln 3}{2}-\frac{2 m q}{3\sqrt 3}\right)+\ldots\right],
\ee
where $...$ denotes subleading terms in the $N$, $\eta$  and $q$ expansions.

Recalling our discussion in Sec. \ref{sec:staticland}, it is instructive to study the analog of  $\#_{av.-min}^{saddle}(inf)$ given by $\#_{av.-min}^{saddle}(inf  \cap min)$ which is the number of inflationary regions per minimum which have a graceful exit, again averaged over all minima:
\begin{eqnarray}
\#_{av.-min}^{saddle}(\mathrm{min}\cap \mathrm{inf})&\equiv&\frac{P(\mathrm{min} \cap \mathrm{inf})}{P(\mathrm{min})}\nonumber\\
&& \nonumber\\
&=&\frac{P(\mathrm{inf})}{P(\mathrm{min})}P(\mathrm{min}| \mathrm{inf})=\#_{av.-min}^{saddle}(\mathrm{inf})P(\mathrm{min}| \mathrm{inf})\quad.
\end{eqnarray}
Using the above results we can write this as
\be
\#_{av.-min}^{saddle}(\mathrm{min}\cap \mathrm{inf})\simeq \exp \left[ -\beta N^2 \left\{ \frac{\ln 3}{4}-\frac{2}{3 \sqrt 3}(\eta + m q)\right\}\right],
\ee
which we compare with
\be\label{eq:SaddleBoost}
\#_{av.-min}^{saddle}(\mathrm{inf})\simeq \exp\left( \beta N^2 \frac{4}{3 \sqrt 3} \eta\right).
\ee
We see that unless $\eta\gtrsim \frac{3 \sqrt 3 \ln 3}{8}\sim 0.7$, the probability of finding a graceful exit after inflation exponentially suppresses the average number of inflationary saddle points per minimum which do have a good exit into a minimum. Fig.~\ref{fig:saddleCount} displays this behaviour for an example where $\eta=0.1$.

Observational constraints, e.g. spatial 3-curvature $|\Omega_k|\lesssim 10^{-3}$, coupled to the standard assumptions about the post-inflationary history \footnote{For well motivated exceptions in the context of string cosmology see e.g. \cite{Dutta:2014tya} and \cite{Cicoli:2016olq}. }, force the total number of e-folds  to be $N_e\gtrsim 60$. It is easy to see that this implies $\eta\lesssim 0.1$ for inflationary saddle points with a sub-Planckian field range. For such saddle points we can write the scalar potential in the vicinity of the critical point (either $V'=0$ -- saddle, or $V''=0$ -- inflection point) in a series expansion as
\be
V=V_0\times \left\{\begin{array}{c}\left(1-\sqrt{2\epsilon_0} \phi -\frac{1}{p\Delta\phi^p}\phi^p+\ldots\right) \\ \\ \left(1-\frac{\eta_0}{2} \phi^2 -\frac{1}{p\Delta\phi^p}\phi^p+\ldots\right)\end{array}\right. \quad,
\ee
where $\phi$ denotes the direction in field space along the solution to the background equations of motion.
Hence, an inflationary critical point is determined by the slow-roll parameters $\epsilon_0$ and $\eta_0$ at the critical point ($\phi=0$) and the width $\Delta\phi$ of the flat region around the critical point, beyond which the potential gets steep quickly.
A sufficiently long slow-roll phase requires $\epsilon_0,\eta_0\ll1$, while inflation ends at $\phi_e\lesssim \Delta\phi$ where $\epsilon=1$. Hence, typical small-field saddle/inflection points have $\phi_0\ll M_{\rm P}$. The expression for the first slow roll parameter $\epsilon$
\be
\sqrt{2\epsilon}\simeq\left\{\begin{array}{c}\sqrt{2\epsilon_0}-\frac{\phi^{p-1}}{\Delta\phi^p} \\ \\ \eta_0\phi-\frac{\phi^{p-1}}{\Delta\phi^p}\end{array}\right.
\ee
implies that the number of e-foldinds $N_e=\int 1/\sqrt{2 \epsilon}\ d \phi$ scales like
\be
N_e(\phi)\sim \frac{\Delta\phi^p}{\phi^{p-2}}\quad.
\ee
Hence, we get an approximate expression for the size of $\eta$ at $N_e$ e-folds before the end of inflation
\be\label{eq:etaExpect}
\eta(N_e)\simeq \eta_0-(p-1)\frac{\phi^{p-2}}{\Delta\phi^p}\sim \frac{1}{N_e}
\ee
as long as $\eta_0\lesssim 1/N_e$. Therefore, due to the observational requirement $N_e\gtrsim 60$ the behaviour shown in Fig.~\ref{fig:saddleCount} for $\eta=0.1$ is representative for the strong exponential suppression of the frequency of observationally viable inflationary saddle points with viable exit relative to minima.

The root cause of the this suppression is the extreme improbability of finding sustained large deviations of the eigenvalue spectrum of a random landscape from the equilibrium distribution. We therefore expect that any inflationary mechanism, which avoids having to reach a minimum at random starting from a given inflationary region of the scalar potential, but instead guarantees the existence of minimum together with inflationary potential patch by construction, will likely dominate the landscape of inflationary models. One rather obvious class of examples are axionic large-field models of inflation \footnote{For recent application of RMT techniques to such systems see \cite{Wang:2015rel}.}. In such cases the origin and structure of the weakly broken axion shift symmetry essentially guarantees analyticity of the axion potential around the points with vanishing axion potential together with a large-field slow-roll region of the scalar potential. This links the existence of a minimum for a graceful exit from inflation together with the existence of the inflationary region itself, resulting in $P(\mathrm{min}| \mathrm{inf})=1$ for large-field models.

The property of getting $P(\mathrm{min}| \mathrm{inf})\lesssim 1$ for large-field inflation is visible even in the DBM model of relaxing away from an inflationary critical point. The requirement of an approximate symmetry protecting the large-field shape of the scalar potential along the inflaton direction effectively boils down to the statement in the DBM picture, that the correlation length of the scalar potential along the inflation trajectory is significantly larger than the field displacement $\Delta\phi$. Hence, large-field inflation can be crudely modeled in DBM by staying in the regime $\Delta\phi\ll \Lambda_h$ which implies that $q=\exp(-\Delta\phi/\Lambda_h)\simeq 1$ stays close to unity. In this regime the rate function is close to zero, automatically implying $P(\mathrm{min}| \mathrm{inf})\lesssim 1$ for large-field models even using the DBM description itself\footnote{We thank Jonathan Frazer for bringing this point to our attention.}.

Evaluating the 'Drake equations' for inflation in the landscape for random small-field saddles vs axionic large-field models discussed in \cite{Pedro:2013nda,Westphal:2012up} using the above results yields
\begin{eqnarray}
\frac{\#_{av.-min}^{saddle}(\mathrm{min}\cap \mathrm{inf})}{\#_{av.-min}^{large-field}(\mathrm{min}\cap \mathrm{inf})}&=&\frac{\#_{av.-min}^{saddle}(\mathrm{inf})}{\#_{av.-min}^{large-field}(\mathrm{inf})}\frac{P^{saddle}(\mathrm{inf} | \mathrm{min})}{P^{large-field}(\mathrm{inf} | \mathrm{min})}\nonumber\\ && \nonumber\\
&=&\frac{P^{saddle}(\mathrm{min}\cap \mathrm{inf})}{\#_{av.-min}^{large-field}(\mathrm{inf})}\sim \frac{e^{-\beta \frac{\ln 3}{2} N^2}}{\#_{av.-min}^{large-field}(\mathrm{inf})} \\
&& \nonumber\\
&\ll & \frac{1}{\#_{av.-min}^{large-field}(\mathrm{inf})}\quad.\nonumber
\end{eqnarray}
The number of axionic large-field regions per minimum $\#_{av.-min}^{large-field}(\mathrm{inf})$ is roughly governed by the dimension of the axionic field space of string compactifications modulo e.g. topological and orientifold existence requirements \cite{Westphal:2012up,Silverstein:2013wua}. Unless these requirements lead to strong exponentially suppression $\#_{av.-min}^{large-field}(\mathrm{inf})\ll 1$, we expect therefore a preponderance of large-field inflation compared to random small-field models in the landscape.

Note, that this bias against accidental small-field inflation in the landscape arises independently from any other bias which cosmological population dynamics may introduce. If Coleman-De Luccia (CDL) tunneling is the dominant process responsible for vacuum transitions in the landscape, then~\cite{Westphal:2012up} has shown that the tunneling dynamics treats small-field and large-field inflation on an even footing, giving rise to a roughly flat prior probability factor over field range from the tunneling dynamics. However, as discussed recently in~\cite{Bachlechner:2016mtp} the CDL bubble domain walls directly connecting different dS vacua may produce instabilities due to moduli runaway from 'over-uplifting' inside the bubble walls. This may prevent the use of the direct CDL process for populating a given sector of the dS landscape. In this case there are Farhi-Guth-Guven-instanton based 'double bubble' configurations which can mediate dS-dS vacuum transitions even in the absence of moduli stablilization preserving CDL bubble domain walls. The tunneling probability for these transitions is then argued in~\cite{Bachlechner:2016mtp} to produce a strong exponential bias towards high-scale and thus large-field inflation. We note, that in such situations the tunneling dynamics thus produces an \emph{independent} exponential bias favoring large-field inflation over accidental small-field inflation in the 'landscape Drake equations' of~\cite{Westphal:2012up}. While being more model-dependent in relying on an assumption about the dominant mode of dS-dS tunneling, this possible tunneling bias for large-field inflation acts in addition to the exit probability driven exponential suppression of accidental small-field inflation in the landscape we found above.

\subsection{Implications for the duration of inflation}\label{sec:measure}

If we look at the expression for the relative small-field saddle point count per minimum, Eq.~\eqref{eq:SaddleBoost}, and our generic expectation for $\eta$ in Eq.~\eqref{eq:etaExpect}, we see that this furnishes us with an intrinsic probability factor 
\be
F_{freq.}(\eta,N)\sim \#_{av.-min}^{saddle}(\mathrm{inf})\simeq \exp\left(\beta\Delta\, N^2\eta\right)
\ee
weighting both the curvature $\eta$ of the saddle points at $N_e$ efolds before the end of inflation and the number of fields $N$ participating in the random small-field saddle point inflation. Beyond that, the structure of the saddle point potential provides relations given in the previous subsection between $\eta$ and the number of efolds $N_e$, as well as between the amount of density perturbations $\delta\rho/\rho$ generated during inflation and the scale of the inflaton potential $V_0$, the width $\Delta\phi$, and the potential curvature $\eta$ at$N_e$ efolds before the end of inflation. Following the analysis of~\cite{Freivogel:2005vv} and assuming a smooth probability distribution $F(V_0,\Delta\phi,\eta)$ for the 'microscopic' parameters $V_0,\Delta\phi,\eta$ with range $[0,1]$, we can compute a measure for the amount of efolds $N_e$
\bea
P(N_e)&\sim & \int\limits_0^1 d V_0\, d\Delta\phi\, d\eta\, \delta\left(N_e-\frac{1}{\eta}\right)\, \delta\left(\frac{\delta\rho}{\rho}-\frac{\sqrt{V_0}}{\eta^{\frac{p-1}{p-2}}\Delta\phi^{\frac{p}{p-2}}}\right)\nonumber\\
&&\hspace{25ex} \times F_{freq.}(\eta,N)\, F(V_0,\Delta\phi,\eta)\\
&&\nonumber\\
&=& \frac{2(\delta\rho/\rho)}{N_e^{4+\frac{2}{p-2}}}\,F_{freq.}(1/N_e,N)\,\int\limits_0^1 d\Delta\phi\, \Delta\phi^{\frac{2p}{p-2}}\,F\left(\frac{\delta\rho^2}{\rho^2}\,\frac{\Delta\phi^{\frac{2p}{p-2}}}{N_e^{2\frac{p-1}{p-2}}},\Delta\phi,\frac{1}{N_e}\right)\;.
\eea
For simplicity, now let us assume one possible variant of neutrality for the prior measure $F(V_0,\Delta\phi,\eta)$ on the microscopic parameter space by choosing a flat prior $F=1$. To argue this, all we need to assume is that $F(\mu_i)$ is a bounded function on the interval of natural values $-1\lesssim \mu_i\lesssim 1$ where the parameters $\mu_i$ take their values. Then $F(\mu_i)\simeq const.$ for $|\mu_i|\ll 1\;\forall\ i$.

Imposing thus $F=1$ the final integral yields
\be
P(N_e)\sim \frac{2(\delta\rho/\rho)}{N_e^{4+\frac{2}{p-2}}}\,F_{freq.}(1/N_e,N)\sim \frac{2(\delta\rho/\rho)}{N_e^{4+\frac{2}{p-2}}}\,e^{{\cal O}(1)\,\frac{N^2}{N_e}}\quad.
\ee

While we find this way that the absolute probability to get more efolds than required by observations $N_e > 62$ is very small, we can now ask for the anthropically conditioned probability that we get $N_e>62$ conditioned on having at least $N_e>59.5$ efolds for structure formation to avoid disruption by too strong negative spatial curvature $\Omega_k$. This conditional probability evaluates to be
\be
P(N_e>62|N_e>59.5)=\frac{\int\limits_{62}^\infty dN_e P(N_e)}{\int\limits_{59.5}^\infty dN_e P(N_e)} \gtrsim 0.9 \quad \Leftrightarrow \quad N\lesssim 10\quad.
\ee
Hence, the probability that $N_e$ is large enough to give observationally viable $\Omega_k$, provided that structure formation was successful ($N_e>59.5$), is about $90\,\%$ \emph{if the number of fields participating in inflation is small $N\lesssim 10$}. Turning this around, the non-observation of sizable negative spatial curvature implies that multi-small-field saddle point inflation models arising at random in the string landscape with more than $N\sim 10$ fields are strongly disfavored by the current bounds on $\Omega_k$. Hence, observationally viable random small-field inflation even in a high-dimensional landscape is driven by just $N\sim a\ few$ fields, which limits the use of robust large-$N$ regime for small-field models in string theory.

\section{Discussion}

In this paper we try to quantify the probability of a graceful exit in the string landscape through the use of random matrix theory techniques. We model the landscape by a Gaussian ensemble, a choice that is simple enough to be solvable and yet whose structure is rich enough to provide the necessary features for a qualitative description of the string landscape. The problem in hand consists in the determination of the probability of connecting a patch in field space where the mass spectrum is slightly tachyonic with another where it is positive definite via a solution to the equations of motion. A typical string theory example for such sectors of the landscape without long-range structures is the scalar potential for the $h^{2,1}\gg 1$ complex structure moduli of a generic non-trivial Calabi-Yau compactification of string theory.

With that particular problem in mind we develop a new method for estimating the transition probabilities in Dyson Brownian motion. It relies on the saddle point evaluation of the partition function and allows for the analytical estimation of transition probabilities between different eigenvalue spectra. The proposed method is exact whenever the initial state matrix is well characterised by a single variable, typically its average eigenvalue, and gives the correct scaling of the transition probability over long "time" scales or equivalently beyond one correlation length. The method is particularly useful whenever one is dealing with transitions to highly unlikely eigenvalue spectra, cases which are computationally very intensive with the traditional techniques. It is clear that this method is not limited to the particular problem we are interested in and therefore we believe that it may find applications in other fields where RMT plays a role.

In order to apply the method to the string landscape we first set the initial conditions to be given by an ensemble of Hessians with an eigenvalue distribution describing slow-roll flat inflationary critical points. We have calculated the probability of a graceful exit from such a random inflationary critical point by applying the saddle point computation results. This led to our central result that the exit probability for small-field inflation in the landscape is exponentially small. The suppression exponent increases quadratically with number of light fields $N$.

We compared this behaviour of small-field inflation in the landscape with large-field models which usually have a viable graceful exit minimum built-in by virtue of the underlying structure and/or symmetry. Taken at face value, this exponentially  disfavours small-field inflation in the landscape. 

Finally, we analyzed the influence of the $\exp(-c N^2)$ suppression of small-field inflation on the probability of observing negative spatial curvature in a landscape where the various dS vacua and inflationary critical points are populated via Coleman-De Luccia (CDL) tunnelling transitions. The exponentially strong dependence on the number of light fields $N$ participating in a small-field inflationary critical point leads to an exponentially strong posterior probability distribution function for $N$ derived from the non-observation of spatial negative curvature. Evaluating this bound for the observed bounds on negative curvature, we found a severe limit on the effective number such light fields $N\ll 10$.

\section*{Acknowledgments}
We thank Ben Freivogel, Enrico Pajer, Liam McAllister, David Marsh, Tim Wrase and Kepa Sousa for interesting discussions.
\vspace{4mm}
\\FGP is supported by the ERC Advanced Grant SPLE under contract ERC-2012-ADG-20120216-320421, by the grant FPA2012-32828 from the MINECO,  and the grant SEV-2012-0249 of the ``Centro de Excelencia Severo Ochoa" Programme. 
The work of AW is supported by the ERC Consolidator Grant STRINGFLATION, ERC-2014-CoG-20152020-647995.

\end{document}